\documentclass[aps,prc,preprint,showpacs,floatfix,superscriptaddress,
nofootinbib]{revtex4}
\usepackage[dvips]{graphicx}
\newcommand{\be}{\begin{equation}} \newcommand{\ee}{\end{equation}}
\newcommand{\ba}{\begin{eqnarray}} \newcommand{\ea}{\end{eqnarray}}
  \newcommand{\nk}{{\bf k}} \newcommand{\nq}{{\bf q}}
\newcommand{\np}{{\bf p}} \newcommand{\nh}{{\bf h}}
\newcommand{\nO}{{\bf 0}} \newcommand{\nkappa}{\mbox{\boldmath
    $\kappa$}} 
\newcommand{\neta}{\mbox{\boldmath $\eta$}}

\begin{document}

\noindent
\title{Inelastic electron-nucleus scattering and scaling 
   at high inelasticity}
%

\bigskip
\bigskip

\author{M.B. Barbaro}
\affiliation{Dipartimento di Fisica Teorica, Universit\`a di Torino
    and INFN, Sezione di Torino, Via P. Giuria 1, 10125 Torino, Italy}
\author{J.A. Caballero}
\affiliation{Departamento de F\'\i sica At\'omica, Molecular y Nuclear, 
 Universidad de Sevilla, Apdo. 1065, E-41080 Sevilla, Spain}
\author{T.W. Donnelly}
\affiliation{Center for Theoretical Physics,
    Laboratory for Nuclear Science and Department of Physics
    Massachusetts Institute of Technology, Cambridge, MA 02139, USA }
\author{C. Maieron}
\affiliation{Departamento de F\'\i sica At\'omica, Molecular y Nuclear, 
 Universidad de Sevilla, Apdo. 1065, E-41080 Sevilla, Spain}
\affiliation{INFN, Sezione di Catania, Via S. Sofia 64, 95123 Catania,
Italy}
\begin{abstract} 

Highly inelastic electron scattering is analyzed within the context of the unified
relativistic approach previously considered in the case of quasielastic kinematics.
Inelastic relativistic Fermi Gas modeling that includes the complete inelastic
spectrum --- resonant, non-resonant and Deep Inelastic Scattering --- is elaborated and
compared with experimental data. A phenomenological extension of  the model based on
direct fits to data is also introduced. Within both models, cross sections and response
functions are evaluated and binding energy effects are analyzed. Finally, 
an investigation of the second-kind scaling behavior is also presented.

\end{abstract} 


\maketitle

\section{Introduction}
\label{sec:intro}

In this work we consider highly inelastic electron
scattering and compare its analysis with the case of quasielastic (QE)
electron scattering. The latter is dominated by the
process where the exchanged virtual photon interacts with a nucleon in
the nuclear ground state and ejects that nucleon, thereby forming a
nuclear particle-hole excitation. 
Corrections to this dominant
process involve going beyond the impulse
approximation to account for two-body currents, 
final-state interactions and nuclear
correlations. Although these contributions
are known not to be entirely negligible 
\cite{Alberico:1989aj,Alberico:1993ur,Barbaro:1995ez,Amaro:1998ta,
Amaro:2001xz,Amaro:2002mj,Amaro:2003yd}
this simple process accounts for the basic feature seen in
the vicinity of elastic scattering from a nucleon at rest, namely, the
QE peak. Models such as those discussed below take into account the
fact that the nucleons in the nucleus are moving and are bound and
thereby produce a broad peak in the inelastic spectrum. In the present
work our goal is to extend the analysis, still maintaining the same
basic features of the relativistic modeling used for the QE region,
and now focus on what we call highly inelastic scattering, or for
brevity, simply the inelastic region. This includes everything that
goes beyond the QE process: that is, whereas the QE process assumes
{\em elastic} scattering from the nucleons, the inelastic process
will assume {\em inelastic} e-N
scattering. For relatively low final-state invariant masses one lies
in the region of resonance excitation and two cases of this sort have
been explored in recent work \cite{Amaro:1999be,Alvarez-Ruso:2003gj}.
In the present study these ideas are generalized to include
the complete inelastic spectrum, both resonant and non-resonant,
including Deep Inelastic Scattering (DIS), within the context of the
unified relativistic approach used in our previous work.
  
Thus, in the present work our goal is to begin by exploring extensions of the 
Relativistic Fermi Gas (RFG) model 
\cite{Serot:1984ey,Alberico:1988bv,Amaro:2002mj}
to an inelastic version of this approach. While this bears some connection with traditional
convolution models for the high-energy response of nuclei
(see for example \cite{Bick,Norton,Simula,BodekRitchie}) 
it is not the same
in that, albeit within a model, it correctly incorporates a specific 
{\it relativistic} nuclear spectral
function into the problem, whereas some other approaches make additional
assumptions and use only the integral of the spectral function, namely, the
nuclear momentum distribution or make non-relativistic approximations
when dealing with the spectral function.

This distinction can be seen quite clearly in
studies of first- and second-kind scaling 
\cite{Alberico:1988bv,Day:1990mf,Donnelly:1998xg,
Barbaro:1998gu,Donnelly:1999sw,Maieron:2001it}
and will not be
elaborated here. Once the inelastic RFG modeling is in hand, it becomes clear
that it might be useful to explore a phenomenological extension of this
model, namely what we call the Extended Relativistic Fermi Gas (ERFG). In this
approach we take the result of doing the correct integral over the nuclear
spectral function (i.e., not the full integral, which is the momentum 
distribution, as alluded
to above) directly from fits made previously to the data~\cite{Jourdan:1996}. 
We shall see that this has a significant impact on the nuclear responses at
high inelasticity.

An issue which will also become clear later is 
that the story is not yet complete:
in addition to the modeling done in the present work, where 
the focus is placed on
incorporating inelastic effects at high energies, there are still 
other contributions
that must be added. Specifically, in recent work \cite{DePace:2003xu} on
2p-2h meson-exchange current effects it is seen that 
a significant incoherent contribution
must be added to those explored here. 
Given that the work on 2p-2h effects is, as yet,
incomplete --- correlation contributions are presently being included --- 
it is 
premature to make too much of comparisons with experimental data, 
and, as we remark later
in the appropriate places, the final understanding of how all of 
the various reaction
mechanisms enter, while becoming clearer is not yet achieved.

The paper is  organized as follows:
in Sec.~\ref{sec:form} we recall the general formalism for inelastic
electron-nucleus scattering;
in Sec.~\ref{sec:RFG} we derive the expressions for the inelastic
hadronic tensor in three different models:
the pure RFG model (\ref{subsec:RFG}), the RFG including the 
effects of binding energy (\ref{subsec:shift}) 
and the ERFG (\ref{subsec:ERFG});
in Sec.~\ref{sec:results} we present numerical results for cross sections
(\ref{subsec:cross}), response functions (\ref{subsec:resp}) and
scaling functions (\ref{subsec:scaling}) 
and, finally, in Sec.~\ref{sec:concl}
we draw our conclusions.

\section{Inelastic electron-nucleus scattering: general formalism}
\label{sec:form}

With the goal outlined above in mind, we start by rewriting the general 
expressions that apply in both elastic and inelastic regimes. 
The general formalism describing inclusive electron-nucleus scattering
processes is widely available~\cite{DeForest:dn,Boffi:gs,Donnellyvar};
here we simply focus on those aspects that are of special relevance for the
discussion that follows. We follow the conventions and 
metric of~\cite{Bjo65} and use capital letters
to refer to four-vectors. The incident and scattered electron
four-momenta are denoted by $K_i^\mu=(\varepsilon_i,\nk_i)$ and
$K_f^\mu=(\varepsilon_f,\nk_f)$.  The hadronic variables, $P_A^\mu=(M_A,\nO)$ 
and $P_B^\mu=(E_B,\np_B)$ represent the
four-momenta of the target and residual nucleus, respectively. The
four-momentum transfer is given by $Q^\mu=(\omega,\nq)$ (we assume the
Born approximation, i.e., only one virtual photon exchanged in the
process).  

Following standard procedures the differential cross section may be written
\be
\frac{d\sigma}{d\Omega_f d\varepsilon_f} = \frac{2\alpha^2}{Q^4}
\frac{\varepsilon_f}{\varepsilon_i} \eta_{\mu\nu} W^{\mu\nu} \, , 
\ee
where $\alpha$ is the fine structure constant, $\eta_{\mu\nu}$ is the
leptonic tensor that can be evaluated directly using trace
techniques~\cite{Donnelly:ry},
and $W^{\mu\nu}$ is the hadronic tensor
containing all of the nuclear structure and dynamics information. 
Assuming that the final state can be described in terms of a recoiling nuclear
state $| \psi_B\rangle$ plus a (highly) inelastic state 
$| \Phi_X\rangle$, its
general expression is given by
\ba
W^{\mu\nu} &=& \overline{\sum_A} \sum_B \sum_X
\langle \psi_B, \Phi_X \left|\hat J^\mu (\nq) \right| \psi_A \rangle^\ast 
\langle
\psi_B, \Phi_X\left|\hat J^\nu (\nq)\right| \psi_A\rangle 
\nonumber \\\
&& \times \rho(E_B) dE_B \rho(E_X) 
dE_X
\delta(\varepsilon_i-\varepsilon_f+E_A-E_B-E_X) \, ,
\label{eq:Wmunu}
\ea 
where $\overline{\sum}_A$ ($\sum_B \sum_X$) indicates the appropriate
average (sum) over initial (final)  states. 
Here $\hat J^\mu (\nq)$ is the Fourier transform of the nuclear current
operator evaluated, $|\psi_A\rangle$ and $|\psi_B,\Phi_X\rangle$ represent
the initial and final states, respectively, and
the distribution functions $\rho(E_B)$
and $\rho(E_X)$ are introduced to account for the
energy-momentum dispersion relation of the final nuclear ($B$) 
and hadronic ($X$) systems.
In this work we assume that the inelasticity of the process is
totally accounted for by the final state $\Phi_{_{X}}$; 
hence for the energy distribution function of the residual
nuclear system we use $\rho(E_B)=\delta(E_B-\overline{E}_B)$, where
$ \overline{E}_B=\sqrt{\np_B^2+ (M^\ast_B)^2}$.
Note that $W^{\mu\nu}$ in Eq.~(\ref{eq:Wmunu}) is meant to be evaluated at
$\np_B + \np_X=\nq=\nk_i-\nk_f$.

The nuclear tensor can equivalently be expressed as an integral in the
$({\cal E},p)$ plane, with $-\np = \np_B$ 
the three-momentum of the recoiling 
daughter nucleus and 
${\cal E}\equiv\sqrt{\np^2+(M^{*}_B)^2}-\sqrt{\np^2+(M^{0}_B)^2}$
the excitation energy of the residual nucleus (see~\cite{Day:1990mf}).
The domain of integration is the kinematically allowed region 

\be
\max[{\cal E}(0),0]\le{\cal E}\le{\cal E}(\pi)
\ee
where
\be
{\cal E}(\theta)=M_A^0+\omega
-\sqrt{\left(M_B^0\right)^2+p^2}-\sqrt{W_{_X}^2+q^2+p^2+2 p q\cos\theta}\,,
\ee
with $\theta$ the angle between $\np$ and $\nq$, and where $W_{_X}$ is
the invariant mass of the final state.
In the $M_B^0\to\infty$ limit the above expression becomes
\be
{\cal E}_\infty(\theta)=m_N+\tilde\omega
-\sqrt{W_{_X}^2+q^2+p^2+2 p q\cos\theta}\,,
\ee
where $m_N$ is the nucleon mass, $\tilde\omega\equiv\omega-E_S$, and $E_S=M_B^0+m_N-M_A^0$ is the 
separation energy.

The upper curve ${\cal E}(\pi)$ crosses the $p$-axis at $p_-=-y_{_X}$
and $p_+=Y_{_X}$, where
\be
y_{_X}=\frac{1}{2W^2}\left[
(M_A^0+\omega)
\sqrt{(W-M_B^0)^2-W_{_X}^2}\sqrt{(W+M_B^0)^2-W_{_X}^2}
-2 q\Lambda_{_X}\right]
\ee
and
\be
Y_{_X}=\frac{1}{2W^2}\left[
(M_A^0+\omega)
\sqrt{(W-M_B^0)^2-W_{_X}^2}\sqrt{(W+M_B^0)^2-W_{_X}^2}
+2 q\Lambda_{_X}\right]\,,
\ee
and where
\be
W=\sqrt{(M_A^0+\omega)^2-q^2}\,\,\mbox{and}\,\,
\Lambda_{_X}=\frac{1}{2}\left[W^2+(M_B^0)^2-W_{_X}^2\right]
\,.
\ee
The variable $y_{_X}$ is the generalization of the usual $y$-scaling variable
to the inelastic process where a resonance $X$ is produced. In the limit
$M_B^0\to\infty$ it reads
\be
y_{_{X,\infty}}=\sqrt{(\tilde\omega+m_N)^2-W_{_X}^2}-q\,.
\ee

Note that the allowed region decreases with $W_{_X}$ and collapses to a point
when $-y_{_X}=Y_{_X}$, which implies $W=M_B^0+W_{_X}$ or, in the $M_B^0\to\infty$ 
limit, $(y_{_{X,\infty}})_{min}=-q$, corresponding to $(W_{_X})_{max}=
\tilde\omega+m_N$. Summarizing, for fixed four-momentum transfer,
the resonant mass is limited to the range
\be
m_N+m_\pi\le W_{_X}\le m_N+\omega-E_S\,.
\label{eq:Wrange}
\ee

\section{The Relativistic Fermi Gas model}
\label{sec:RFG}

In this section we proceed by evaluating the hadronic nuclear tensor
assuming the impulse approximation and by working within the framework of
the RFG model. In this case, the virtual
photon is absorbed by an on-shell nucleon described by a Dirac spinor
$u(\nh,s_h)$, with energy $\overline{E}_\nh=\sqrt{\nh^2+m_N^2}$.
Integrating over the momenta in the Fermi sea,
the following expression for the inelastic hadronic tensor results 
\ba
W^{\mu\nu}(\nq,\omega) 
&=&\frac{3{\cal N}}{4\pi p_F^3} \int_F
d\nh\frac{m_N}{\overline E_\nh} \int dE_{_X} \delta(\omega+{\overline
  E_\nh}-E_{_X})
\nonumber \\
&\times & \frac{1}{2} \sum_{s_h} \sum_{X_i} \rho(E_{_{X_i}})
\left[\overline \Phi_{_{X_i}} \hat J^\mu u(\nh,s_h)\right]^\ast
\left[\overline \Phi_{_{X_i}} \hat J^\nu u(\nh,s_h)\right] \, ,
\label{eq:WRFG}
\ea where ${\cal N}$ is the number of nucleons (protons or neutrons)
and $\int_F d\nh\equiv\int d\nh\theta(p_F-h)$, $p_F$ being the Fermi
momentum.  
The symbol $\int dE_{_X}$ stands for the integral
over the energy of the inelastic final state, while $\sum_{X_i}$
indicates in general the sum/integral over all the internal quantum
numbers of all possible inelastic final states $\Phi_{_{X_i}}$,
having total energy $E_X$ and total momentum $\np_X$, fixed by momentum
conservation to be $\np_X = \nh + \nq$. 

The hadronic tensor in Eq.~(\ref{eq:WRFG}) for inelastic processes can be
also written in the form 
\be W^{\mu\nu}_{inel}(\nq,\omega) =\frac{3{\cal N}}{4\pi
  p_F^3} \int dE_{_X} \int_F d\nh\frac{m_N}{{\overline E_\nh}}
w^{\mu\nu}_{inel}(H,Q,E_{_X}) \delta(\omega+{\overline E_\nh}-E_{_X})
\, ,
\label{eq:Winel}
\ee 
where 
$H^\mu=(\overline{E}_\nh,\nh)$
and we have introduced the inelastic single-nucleon tensor 
\be
w^{\mu\nu}_{inel}(H,Q,E_{_X})=\frac{1}{2} \sum_{s_h} \sum_{X_i}
\rho(E_{_{X_i}}) \left[\overline \Phi_{_{X_i}} \hat J^\mu
  u(\nh,s_h)\right]^\ast \left[\overline \Phi_{_{X_i}} \hat J^\nu
  u(\nh,s_h)\right] \, .  
\ee 
Note that the above single-nucleon
tensor has dimensions of $E^{-1}$. As will be shown later, this is
in contrast with our past work on QE and $N\to\Delta$ scattering where the single-nucleon tensors were defined to be dimensionless.

Next we choose to express the inelastic hadronic tensor in Eq.~(\ref{eq:Winel})
in terms of the invariant mass $W_{_X}$, 
\be 
W^{\mu\nu}_{inel}(\nq,\omega) =\frac{3{\cal N}}{4\pi
  p_F^3} \int dW_{_X} \int_F d\nh\frac{m_N W_{_X}}{{\overline E_\nh} E_{_X}}
w^{\mu\nu}_{inel}(H,Q,E_{_X}) \delta(\omega+{\overline E_\nh}-E_{_X})
\label{eq:Winel1}
\ee 
with $E_{_X}=\sqrt{\np_{_X}^2+W_{_X}^2}$. The energy integral can be
performed by exploiting the delta-function, yielding 
\be
W^{\mu\nu}_{inel}(\nq,\omega) =\frac{3{\cal N}}{4\pi p_F^3} \int_F
d\nh\frac{m_N}{{\overline E_\nh}}
w^{\mu\nu}_{inel}(H,Q,\omega+\overline{E}_\nh) \ .
\label{eq:Winel2}
\ee 
In the case of DIS on a single
nucleon, the inelastic tensor simply reduces to the single-nucleon
tensor $w^{\mu\nu}_{inel}$.

Before entering into a detailed analysis of the inelastic nuclear
tensor, it is interesting to notice how the usual expressions for the
QE and $N\to\Delta$ hadronic tensors are recovered from the
general result given in Eq.~(\ref{eq:WRFG}). First, in the case of
QE scattering, the nuclear final state is simply a
particle-hole excitation, hence, in the RFG model, $\Phi_{_X}$
describes an on-shell nucleon, namely,
$\Phi_{_X}=\sqrt{m_N/{\overline E_\np}}u(\np,s_p)$. The energy
distribution function is simply $\rho(E_{_X})=\delta(E_{_X}-{\overline
  E_\np})$ and the sum over the final states reduces to a sum over spin projections, $\sum_{X_i}=\sum_{s_p}$.  The QE hadronic tensor then reads 
\be 
W^{\mu\nu}_{QE}(\nq,\omega) =\frac{3{\cal N}}{4\pi p_F^3} \int_F
d\nh\frac{m_N^2}{{\overline E_\nh}{\overline E_\np}}
w^{\mu\nu}_{QE}(H,Q) \delta(\omega+{\overline E_\nh}-{\overline
  E_\np}) \, , 
\ee 
where $w^{\mu\nu}_{QE}$ is the usual dimensionless
QE single-nucleon tensor
\be
w^{\mu\nu}_{QE}=\frac{1}{2} \sum_{s_h} \sum_{s_p} \left[\overline
  u(\np,s_p) \hat J^\mu u(\nh,s_h)\right]^\ast \left[\overline
  u(\np,s_p) \hat J^\nu u(\nh,s_h)\right] \, .  
\ee 
In the case of the
transition $N\rightarrow \Delta$, the final state $\Phi_{_X}$, within
the context of the RFG model, is an on-shell Delta, namely,
$\Phi_{_X}=\sqrt{m_\Delta/{\overline
    E_\Delta}}u_\Delta(\np,s_\Delta)$, with on-shell energy
${\overline E_\Delta}=\sqrt{\np^2+m_\Delta^2}$. The energy
distribution function in this case is
$\rho(E_{_X})=\delta(E_{_X}-{\overline E_\Delta})$ and
$\sum_{X_i}=\sum_{s_\Delta}$. The $N\to\Delta$ hadronic tensor that results is
\be 
W^{\mu\nu}_{\Delta}(\nq,\omega) =\frac{3{\cal N}}{4\pi p_F^3} \int_F
d\nh\frac{m_N^2}{{\overline E_\nh}{\overline E_\Delta}}
w^{\mu\nu}_{\Delta}(H,Q) \delta(\omega+{\overline E_\nh}-{\overline
  E_\Delta}) 
\ee 
with $w^{\mu\nu}_{\Delta}$ the dimensionless
nucleon-$\Delta$ tensor 
\be w^{\mu\nu}_{\Delta}=\frac{m_\Delta}{2m_N}
\sum_{s_h} \sum_{s_\Delta} \left[\overline u_\Delta(\np,s_p) \hat
  J^\mu u(\nh,s_h)\right]^\ast \left[\overline u_\Delta(\np,s_p) \hat
  J^\nu u(\nh,s_h)\right] \, .  
\ee 
As expected, these expressions for the
dimensionless single-nucleon tensors coincide with the ones introduced
in~\cite{Donnelly:1991qy,Amaro:1999be}.
Likewise for the Roper resonance the expressions obtained 
in~\cite{Alvarez-Ruso:2003gj} are recovered.

\subsection{The RFG inelastic nuclear tensor and response functions}
\label{subsec:RFG}

In this section we evaluate the inelastic nuclear tensor in the RFG
framework.  For convenience, as usual we first define the dimensionless variables
\ba &&
\kappa^\mu=(\lambda,\nkappa)=\left(\frac{\omega}{2m_N},\frac{\nq}{2m_N}\right),
\,\,\,\, \tau=\nkappa^2-\lambda^2,\,\,\,\,
\eta_F=\frac{p_F}{m_N},\,\,\,\, \epsilon_F=\sqrt{1+\eta_F^2}
\nonumber\\
&& \eta^\mu=({\overline\epsilon},\neta)=\left(\frac{{\overline
      E_\nh}}{m_N}, \frac{\nh}{m_N}\right),\,\,\,\,
\mu_{_X}=\frac{W_{_X}}{m_N},\,\,\,\,
\epsilon_{_X}=\sqrt{\mu_{_X}^2+(\neta+2\nkappa)^2} \,,
\ea
in terms of which the hadronic tensor in Eq.~(\ref{eq:Winel1}) reads
\be W^{\mu\nu}_{inel}(\kappa,\lambda) =\frac{3{\cal N}}{4\pi \eta_F^3} 
\int d\mu_{_X}
\int d\neta\frac{\mu_{_X}}{{\overline\epsilon} \epsilon_{_X}}
w^{\mu\nu}_{inel}(\neta,\mu_{_X};\kappa,\lambda)
\delta(2\lambda+{\overline\epsilon}-\epsilon_{_X}) \theta(\eta_F-\eta)\ 
.
\label{eq:Wadim}
\ee

Before presenting the explicit results for the RFG response functions,
let us discuss an important ingredient of the calculation,
the single-nucleon inelastic hadronic tensor
$w^{\mu\nu}_{inel}$. For unpolarized scattering,
the latter can be parameterized in terms of
two structure functions, $w_1$ and $w_2$,
according to
\ba 
w^{\mu\nu}_{inel}=
-w_1 \left(g^{\mu\nu}+\frac{\kappa^\mu\kappa^\nu}{\tau}\right)
+w_2 \left(\eta^\mu+\kappa^\mu\rho\right)
\left(\eta^\nu+\kappa^\nu\rho\right)\,.
\label{eq:Wadim3}
\ea
For on-shell nucleons, the structure functions $w_1$ and $w_2$
depend on two variables, the four momentum transfer $Q^2$
and the invariant mass $W_X$ of the final state reached
by the nucleon, or, equivalently, the single-nucleon
Bjorken variable   
\be
x=\frac{|Q^2|}{2H\cdot Q}=\frac{|Q^2|}{W_X^2-m_N^2-Q^2}=
\frac{\tau}{\eta\cdot \kappa}\,.
\label{eq:x}
\ee

In our formalism it is convenient to introduce the inelasticity parameter
~\cite{Amaro:1999be,Alv01} 
\be
\rho\equiv 1+\frac{1}{4\tau}(\mu_{_{_X}}^2-1)\,,
\label{eq:rho}
\ee
the value unity corresponding
to elastic scattering.
Note that $\rho$ is simply linked
to the Bjorken scaling variable of the on-shell nucleon moving 
inside the target nucleus 
by the relation $\rho=1/x$, thus in the following
we will use $\rho$ as argument
of the structure functions $w_1, w_2$.

In presenting our results we will also use
the ``laboratory'' Bjorken variable 
\be 
x_L=\frac{|Q^2|}{2 m_N \omega} =\frac{\tau}{\lambda}\ , 
\ee
corresponding to a single nucleon at rest in the laboratory
frame.  

Let us now return to the inelastic nuclear tensor of Eq.~(\ref{eq:Wadim}):
after performing the polar angular integration by means of the
energy-conserving delta-function one gets
\ba W^{\mu\nu}_{inel}(\kappa,\lambda) &=&\frac{3{\cal N}\tau}{2 \eta_F^3\kappa}
\int_0^{2\pi}\frac{d\Phi}{2\pi}
\int_{\rho_1(\kappa,\lambda)}^{\rho_2(\kappa,\lambda)} d\rho
\int_{\epsilon_0(\rho)}^{\epsilon_F} d{\overline\epsilon}
w^{\mu\nu}_{inel}({\overline\epsilon},\theta_0,\rho;\kappa,\lambda)
\label{eq:Wadim1}
\ea
where
\be \cos\theta_0=
\frac{1}{\kappa\eta}\left(\lambda{\overline\epsilon}-\tau\rho \right)\ 
.
\label{eq:costeta}
\ee
The condition $|\cos\theta_0|\leq 1$ fixes the integration limits over
${\overline\epsilon}$:
\be {\overline\epsilon}\geq\epsilon_0(\rho)\equiv
\kappa\sqrt{\frac{1}{\tau}+ \rho^2}-\lambda\rho\ .
\label{eq:eps_0}
\ee

Moreover, by requiring that $\epsilon_0(\rho)\leq\epsilon_F$ and that
the resonance mass is above the pion-production threshold (i.e.,
$\mu_X\geq \mu_{thresh}\equiv 1+\mu_\pi$) the following region is obtained 
for the integration over $\rho$:
\be 
\left[\rho_1(\kappa,\lambda),\rho_2(\kappa,\lambda)\right] =
\left[  \max\left\{\frac{\lambda\epsilon_F-\kappa\eta_F}{\tau},
    \rho_{thresh}\right\},
\frac{\lambda\epsilon_F+\kappa\eta_F}{\tau}
\right] \,,
\ee
with
\be
\rho_{thresh}=1+\frac{\mu_\pi(\mu_\pi+2)}{4\tau}\,.
\label{eq:rho_thresh}
\ee

Note that the upper integration limit $\rho_2(\kappa,\lambda)$ always
lies below the cutoff corresponding to Eq.~(\ref{eq:Wrange}). Indeed using
Eqs.~(\ref{eq:Wrange}) and (\ref{eq:rho}) and keeping in mind that the 
(negative) separation energy of the RFG is
$E_S^{RFG}=-T_F\equiv m_N(1-\epsilon_F)$, the maximum $\rho$ allowed by
Eq.~(\ref{eq:Wrange}) reads 
\be
\rho_{max}^{RFG}
=1+\frac{1}{4\tau}\left[\left(2\lambda+\epsilon_F\right)^2-1\right]
=\rho_2(\kappa,\lambda)+\frac{1}{4\tau}\left(2\kappa-\eta_F\right)^2\,,
\ee
$\rho_2$ therefore resulting in the more stringent integration limit.

Now by writing the single-nucleon inelastic tensor in terms of
structure functions $w_1$ and $w_2$ as in Eq.~(\ref{eq:Wadim3})
and choosing the $z$-direction along $\nq$, the integration over
$\Phi$ and ${\overline\epsilon}$ can be performed analytically (see
Appendix~\ref{app:RFG}) and the hadronic inelastic tensor can be
expressed in the general form
\ba W^{\mu\nu}_{inel}(\kappa,\lambda) &=&\frac{3{\cal N}\tau}{2 \eta_F^3\kappa}\xi_F
\int_{\rho_1(\kappa,\lambda)}^{\rho_2(\kappa,\lambda)}
d\rho
\left(1-\psi_{_X}^2\right)\theta\left(1-\psi_{_X}^2\right)
U^{\mu\nu}(\kappa,\tau,\rho)\,,
\label{eq:Wadim4}
\ea
where $\xi_F=\epsilon_F-1$ is the Fermi kinetic energy and the
inelastic scaling variable
\be \psi_{_X}\equiv \mbox{sign}(\lambda -\tau\rho)
\sqrt{\frac{\epsilon_0(\rho)-1}{\epsilon_F-1}}
\label{eq:psix2}
\ee
has been defined. For each value of $\rho$ (and hence $\mu_X$) a ``peak'' can thus be
identified, corresponding to the region 
$-1\leq\psi_{_X}\leq 1$, 
centered at
\be \psi_{_X}=0\ , \ \ \ 
\lambda_P=\tau_P\rho=\frac{1}{2\rho}\left(\sqrt{1+4\kappa_P^2\rho^2}-1\right)\ , \ \ \ 
\kappa_P=\sqrt{\tau_P\left(1+\tau_P\rho^2\right)}\ ,
\ee
whose width
\be \Delta\lambda =\frac{1}{2}\left[
  \sqrt{\left(2\kappa+\eta_F\right)^2+\mu_X^2}-
  \sqrt{\left(2\kappa-\eta_F\right)^2+\mu_X^2}\right] \simeq
\frac{2\kappa\eta_F}{\sqrt{4\kappa^2+\mu_X^2}} \ee
is a function that grows with $\kappa$ and decreases with $\mu_X$.

The general expression for the tensor $U^{\mu\nu}$ is derived in
Appendix~\ref{app:RFG}. Here we only report the longitudinal and
transverse components
\ba U^L&=&U^{00}=
\frac{\kappa^2}{\tau}\left[\left(1+\tau\rho^2\right)w_2(\tau,\rho)-
  w_1(\tau,\rho)+w_2(\tau,\rho){\cal D}(\kappa,\tau,\rho)\right]
\label{eq:ul}
\\
U^T&=&U^{11}+U^{22}= 2 w_1(\tau,\rho)+w_2(\tau,\rho){\cal D}(\kappa,\tau,\rho)
\label{eq:ut}
\ea
which are linked to the longitudinal and transverse response functions by the
following relations:
\ba R^{L,T}_{inel}(\kappa,\tau)
&=&\frac{3{\cal N}\tau}{2
  \eta_F^3\kappa} \xi_F
\int_{-1}^{\psi_X^{\max}(\kappa,\lambda)}
d\psi_X \left|\frac{\partial \rho}{\partial\psi_X}\right|
\left(1-\psi_{_X}^2\right) U^{L,T}(\kappa,\tau,\rho(\psi_X))\ ,
\nonumber\\
\label{eq:rlt}
\ea
with
\be 
\psi_X^{\max}(\kappa,\lambda)=\min\left\{
1,
\sqrt{
\frac{
\kappa\sqrt{\frac{1}{\tau}+ \rho_{thresh}^2}-\lambda\rho_{thresh}-1
}{\xi_F}}
\right\}
\label{eq:psix2_1}
\ee
and
\be
\frac{\partial \rho}{\partial\psi_X}=
-\sqrt{2\xi_F}\frac{\kappa\left(1+\xi_F\psi_X^2\right)
-\lambda\psi_X\sqrt{2\xi_F\left(1+\frac{1}{2}\xi_F\psi_X^2\right)}}
{\tau\sqrt{1+\frac{1}{2}\xi_F\psi_X^2}}\,.
\ee

In Eqs.~(\ref{eq:ul},\ref{eq:ut}) the function
\ba {\cal D}(\kappa,\tau,\rho) &=& \frac{1}{\epsilon_F-\epsilon_0(\rho)}
\int_{\epsilon_0(\rho)}^{\epsilon_F}d\overline\epsilon \int_0^{2\pi}
\frac{d\Phi}{2\pi} \left( \neta\times\hat{\nkappa}\right)^2
\nonumber\\
&=&
\frac{\tau}{\kappa^2}
\left\{
\frac{1}{3}\left[\epsilon_F^2+\epsilon_F\epsilon_0(\rho)+\epsilon_0(\rho)^2\right]
+\lambda\left[\epsilon_F+\epsilon_0(\rho)\right]
+\lambda^2\right\}-(1+\tau)
\nonumber\\
&+&(\rho-1)\frac{\tau}{\kappa^2}
\left\{\lambda\left[\epsilon_F+\epsilon_0(\rho)\right]
-\tau(\rho+1)\right\}
\nonumber\\
&=&
\xi_F\left(1-\psi_X^2\right) 
\left[
1+\xi_F\psi_X^2
-\frac{\lambda}{\kappa}\psi_X
\sqrt{\xi_F\left(2+\xi_F\psi_X^2\right)}
+\frac{\tau}{3\kappa^2}
  \xi_F\left(1-\psi_X^2\right) 
\right]
\nonumber\\
\label{eq:cald}
\ea
arises from the Fermi motion and goes to zero as $\xi_F\to 0$; being proportional
to $\xi_F\cong \eta_F^2/2\ll 1$, this provides relatively moderate 
corrections to the rest of the contributions in Eqs.~(\ref{eq:ul}) and (\ref{eq:ut}).

The value $\rho=1$ corresponds to QE kinematics: in this
case the well-known expressions for the QE responses
are recovered. The ``total'' observables are then obtained by adding
the usual RFG QE response to the inelastic results:
\be R_{tot}^{L,T} = R_{QE}^{L,T} + R_{inel}^{L,T} \,.
\label{eq:totalresp}
\ee

In the deep inelastic regime it is customary to deal with nuclear
structure functions $W_{1,2}^A$ and/or $F_{1,2}^A$.  These can be
expressed in terms of the longitudinal and transverse response functions
through the following relations
\ba W_1^A &=& \frac{1}{2} R^T
\label{eq:W1A}
\\
W_2^A &=& \left(\frac{\tau}{\kappa^2}\right)^2 R^L +
\frac{1}{2}\frac{\tau}{\kappa^2}R^T
\label{eq:W2A}
\ea
and
\ba F_1^A &=& m_N W_1^A
\label{eq:F1A}
\\
F_2^A &=& 2 m_N \lambda W_2^A \,.
\label{eq:F2A}
\ea

\subsection{Effects of binding energy}
\label{subsec:shift}

In the study of superscaling for inclusive QE electron
scattering from nuclei, an appropriate scaling variable $\psi^\prime$
was introduced by including a small energy shift to have the
QE peak occur at the place where the scaling variable is
zero. A detailed study of the sensitivity of the scaling function to
variations of the Fermi momentum and energy shift was presented 
in~\cite{Donnelly:1998xg,Donnelly:1999sw,Maieron:2001it}. Here we extend this
analysis to the inelastic region. In principle, the introduction of an
energy shift, $\omega_{shift}$, in the formalism is straightforward
and the calculation of the inelastic responses proceeds as in the
$\omega_{shift}=0$ case. However, as will be made clear in the
following, some complications arise. First, due to the general form
assumed for the single-nucleon inelastic hadronic tensor, a certain
asymmetry appears between the energy shift effects in the longitudinal
and transverse responses. These shift effects are larger in the
longitudinal response. Notice that this asymmetry already enters at
the level of the QE nuclear responses. Second, there exists
an ambiguity in the definition of the variable which should be used as
the Bjorken $x$-scaling variable corresponding to the moving nucleon.

The effects of the inclusion of an energy shift on the inelastic
nuclear hadronic tensor have been studied in the literature,
with particular emphasis on the structure function $F_2$ and the
EMC effect at large values of $x$, 
in the context of so called ``binding models''
(see for example \cite{Sick, Ciofi} and the general reviews
\cite{Bick, Norton}). The approach we follow here is the self-consistent
generalization of previous works  on the RFG.
It is formally similar to the binding model approach, where in general
the on-shell energy of the initial nucleon is modified by subtracting
a constant term which effectively accounts for the nucleon separation energy and
for the possibility that the residual nuclear system is left in an (highly)
excited state. 
However, since the existing models either focus only on EMC ratios and/or
use more realistic, although generally non-relativistic
wave functions, a precise quantitative comparison with those models
is not possible. As we will discuss in the results sections, our
calculations still miss some ingredients, coming from meson-exchange
currents, and this makes a detailed quantitative comparison with
experimental data premature; it is clear that, when this comparison
will be made in the future, a more in-depth study of binding effects will also 
be needed.  

In the RFG the energy shift is usually introduced by modifying the argument
of the delta-function appearing in the general expression of the
inelastic hadronic tensor in Eq.~(\ref{eq:Winel}), according to $\omega +
\overline{E}_{\nh} - E_X \rightarrow \omega^\prime +
\overline{E}_{\nh} - E_X$, where $\omega^\prime = \omega -
\omega_{shift}$.  Then, introducing the invariant mass
$W_{_X}^{\prime^2}\equiv E_X^2 - \np_{_X}^2 = (\omega^\prime
+\overline{E}_{\nh} )^2 -\np_{_X}^2$, we can write the inelastic
hadronic tensor in the form
\be 
W^{\mu\nu}_{inel}(\kappa,\lambda) =\frac{3{\mathcal N}}{4\pi \eta_F^3} \int
d\mu_{_X}^\prime \int
d\neta\frac{\mu_{_X}^\prime}{{\overline\epsilon} \epsilon_{_X}}
w^{\mu\nu}_{inel}(\neta,\mu_{_X}^\prime;\kappa,\lambda)
\delta(2\lambda^\prime+{\overline\epsilon}-\epsilon_{_X})
\theta(\eta_F-\eta)\;,
\label{eq:wadim_shift}
\ee
where $\mu_X^\prime = W_{_X}^\prime/m_N$ and $\lambda^\prime =
\omega^\prime/(2 m_N)$. As in the unshifted analysis, the delta-function
can be used to perform the polar angular integration, leading to the
result
\be W^{\mu\nu}_{inel}(\kappa,\lambda) = \frac{3{\mathcal N}\tau^\prime}{2
  \eta_F^3\kappa}
\int_{\rho^\prime_1(\kappa,\lambda^\prime)}^{\rho^\prime_2(\kappa,\lambda^\prime)}
d\rho^\prime \int_0^{2\pi}\frac{d\Phi}{2\pi}
\int_{\epsilon_0^\prime (\rho^\prime)}^{\epsilon_F} d{\overline\epsilon}
w^{\mu\nu}_{inel}({\overline\epsilon},\rho^\prime;\kappa,\lambda)
\;,
\label{eq:wadim_shift1}
\ee
where the variable $\rho^\prime$ is defined as
\be 
\rho^\prime
\equiv \frac{2H\cdot Q^\prime}{|Q^{\prime 2}|}
=\left[ 1 +
  \frac{1}{4 \tau^\prime}(\mu_X^{\prime^2} -1)\right]
\label{eq:rhoxprime}
\ee 
and $\tau^\prime \equiv \kappa^2 - \lambda^{\prime^2}$.

The inclusion of the energy shift modifies the integration limits over
$\overline\epsilon$ in the following way:
\be 
{\overline\epsilon}\geq \epsilon_0^\prime(\rho^\prime) \equiv
\kappa\sqrt{\frac{1}{\tau^\prime}+
  \rho^{\prime^2}}-\lambda^\prime\rho^\prime\ .
\label{eq:eps0prime}
\ee
Correspondingly, the region for the integration over $\rho^\prime$ is
given by
\be 
\left[\rho^\prime_1(\kappa,\lambda^\prime),\rho^\prime_2(\kappa,\lambda^\prime)\right]=
\left[{\mathrm{max}} \left\{
  \frac{\lambda^\prime\epsilon_F - \kappa\eta_F}{\tau^\prime},
  1+\frac{\mu_\pi}{4\tau^\prime} (2+\mu_\pi)\right\},
\frac{\lambda^\prime\epsilon_F+\kappa\eta_F} {\tau^\prime}
\right]\, .
\label{eq:rhoxlimprime}
\ee

The definition of the inelastic scaling variable becomes now
\be \psi_{_X}^{\prime^2} \equiv
\frac{\epsilon_0^\prime(\rho^\prime)-1}{\epsilon_F-1}\,
\label{eq:psix2prime}
\ee and the inelastic longitudinal and transverse response functions,
calculated as $R^L_{inel}=W^{00}_{inel}$ and
$R^T_{inel}=W^{11}_{inel}+W^{22}_{inel}$, have the following general
forms:
\be R^{L,T}_{inel}(\kappa,\tau)= \frac{3{\mathcal{N}}\tau^\prime} {2
  \eta_F^3\kappa} \xi_F \int_{\rho^\prime_1(\kappa,\lambda^\prime)}
^{\rho^\prime_2(\kappa,\lambda^\prime)} d \rho^\prime
 \left(1-\psi_{_X}^{\prime^2}\right)
U^{L,T}(\kappa,\tau,\rho^\prime) \, .
\label{eq:rlt_shift}
\ee In order to evaluate the longitudinal and transverse
nuclear functions $U^{L,T}(\kappa,\tau,\rho^\prime)$  one needs to
assume a specific form for the inelastic single-nucleon tensor
$w^{\mu\nu}_{inel}({\overline\epsilon},\rho^\prime;\kappa,\lambda)$. It
is important to remark that there exists some ambiguity in the
choice made here: for instance, several alternatives involving different
expressions containing the four-momenta $Q^\mu$ and/or $Q^{\prime\mu}$
are possible, and these can lead to different results.
For example, in~\cite{Sick}, the modified four momentum transfer
$Q^{\prime\mu}$ is used, although then a prescription must be used
in order to recover the gauge invariance of the nuclear hadronic tensor,
which is lost by making this choice.
In
Appendix~\ref{app:shift} we present the specific expressions of the inelastic
and QE responses obtained for a given selection of the
single-nucleon tensors accounting for the energy shift. Apart from the
specific form of the tensor $w^{\mu\nu}$, the choice of the
arguments of the single-nucleon inelastic structure functions,
$w_1,w_2$, also present some ambiguities. 
In fact, the available
parameterizations for $w_1, w_2$ that we employ in our calculations
are given for free, on-shell, nucleons, 
while the inclusion of the energy
shift effectively introduces some ``off-shellness'' of the initial
nucleon, by altering the energy balance at the 
vertex where it couples to the exchanged virtual photon.  
In this case the bound-nucleon Bjorken variable is not 
uniquely determined by the final-state invariant mass and,
since no theoretically derived prescriptions exist, one has to
make some assumptions. 
As shown in
Appendix~\ref{app:shift}, the inelasticity parameter selected in this work,
$\tilde{\rho}$, corresponds to the one given as
$(2m_N\tilde{\omega})/|Q^2|= \rho^\prime 
\displaystyle{\frac{\tau^\prime}{\tau}}$, 
where $\tilde{\omega}$ is the energy
transferred to the nucleon in the system in which the nucleon is at
rest. 
This means that in our numerical calculations, for a given set of values
of $\omega$, $Q^2$ and $\rho'$ we employ free-nucleon structure
functions taken at four-momentum $Q^2$ and Bjorken variable $1/\tilde{\rho}$.
 
\subsection{Extended Relativistic Fermi Gas (ERFG)}
\label{subsec:ERFG}

As discussed in previous works \cite{Donnelly:1998xg,Donnelly:1999sw,Maieron:2001it,Amaro:1999be},
for a fixed value of the invariant mass $\mu_X$, the RFG yields
a scaling function 

\be
f(\psi_X^\prime) = f_L(\psi_X^\prime) =  f_T(\psi_X^\prime) =
\frac{3}{4} (1 - \psi_X^{\prime^2}) \theta (1- \psi_X^{\prime^2})
\label{eq:fscaltot}
\ee
which, as a function of the appropriate scaling variable $\psi_X^\prime$,
is the same for all values of $\mu_X$\footnote{The function 
in Eq.~(\ref{eq:fscaltot}) differs from the one used in previous
work~\cite{Maieron:2001it}
by a multiplicative function $\frac{2 \xi_F}{\eta_F^2}
\left[1 +\frac{1}{2}\xi_F\left(1 + \psi_X^{\prime^2}
\right)\right]$. We have checked that this is 
numerically unimportant for all of the kinematical conditions considered here.}.

\begin{figure}[p]
\includegraphics{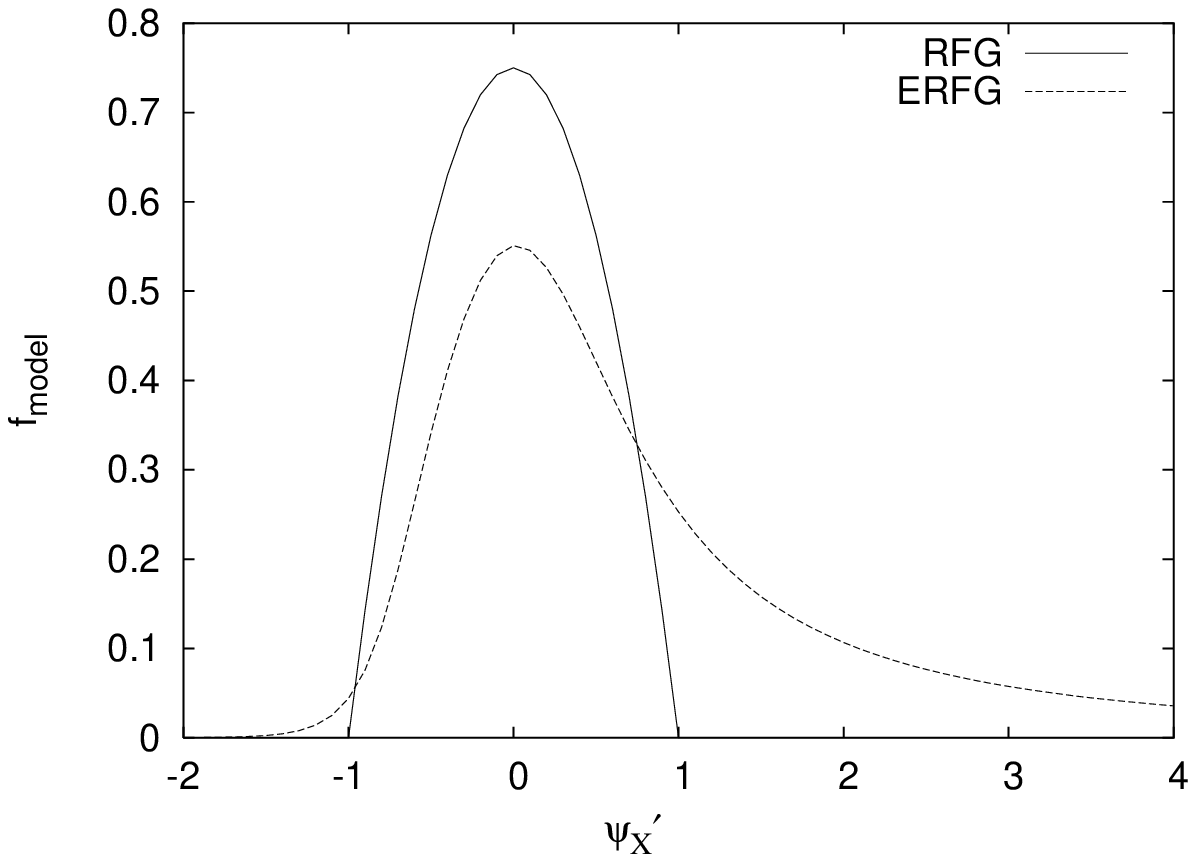}
\caption{
Scaling function $f_{model}(\psi_X')$ of Eq.~(\ref{eq:fmodel})
for the RFG and ERFG models.}
\label{fig:fmodel}
\end{figure}

In~\cite{Donnelly:1999sw} the behavior of the 
longitudinal scaling function 
was studied for the existing world data in the QE region. 
This study showed that to a good approximation $f_L(\psi^\prime)$ 
superscales, that is, it does not show any significant
dependence on the momentum transfer $\kappa$ (scaling of the first kind) 
{\em and} is approximately the same
for all nuclear species (scaling of the second kind). An expression for a phenomenological
longitudinal scaling function, $f_{univ}(\psi^\prime)$, 
was obtained by fitting the data~\cite{Jourdan:1996}.
Based on these results, we now make the following hypothesis:
we assume that  
this $f_{univ}(\psi^\prime)$, derived from the data, provides
a good description of
$f(\psi_X^\prime) = f_L(\psi_X^\prime) =  f_T(\psi_X^\prime)$,
(``scaling of the zeroth kind'') as it implicitly contains the initial-state physics, 
and thus we make, for any $\mu_X$, the following substitution:
\be
\frac{3}{4}
\left( 1- \psi_X^{\prime^2}\right)
\theta \left(1- \psi_X^{\prime^2}\right)
\rightarrow f_{ERFG}(\psi_X^\prime) =f_{univ}(\psi_X^\prime)
\;.
\label{eq:f_univ2}
\ee
To be more specific, we calculate the response functions as
\ba
R_{QE}^{L,T} &=&
\frac{{\mathcal{N}}}
{\eta_F^3\kappa m_N} 
\xi_F f_{model}(\psi^\prime) 
U_{QE}^{L,T} 
\label{eq:rlt_qe_funiv} 
\\ 
R^{L,T}_{inel}(\kappa,\tau) &=& \frac{{\cal N}}{ \eta_F^3\kappa}
\xi_F \int_{\mu_{thresh}}^{1+2\lambda-\epsilon_S} d\mu_X \mu_X
f_{model}(\psi_X^\prime) U^{L,T}\;,
\label{eq:rlt_funiv}
\ea
where $\epsilon_S=E_S/m_N$ is the dimensionless separation energy and
\be
f_{model}(\psi_X^\prime) = 
\left\{
\begin{array}{cc}
\displaystyle{\frac{3}{4}\left(1 - \psi_X^{\prime^2}\right)
\theta\left(1 - \psi_X^{\prime^2}\right)
}
& \qquad {\mathrm{model = RFG}}
\\
f_{univ}(\psi_X^\prime)
& \qquad {\mathrm{model = ERFG}} 
\end{array}
\right.
\label{eq:fmodel}
\ee
The functions $f_{RFG}$ and $f_{ERFG}$ are shown in Fig.~\ref{fig:fmodel} 
as functions of $\psi_X'$, while the functions $U^{L,T}_{QE}$ and $U^{L,T}$ in 
Eqs.~(\ref{eq:rlt_qe_funiv}--\ref{eq:rlt_funiv}) are given 
in Appendix~\ref{app:shift}.

\section{Results}
\label{sec:results}

In this section we present our results for cross sections and
response and structure functions. 
In computing the inelastic hadronic tensor of Eq.~(\ref{eq:wadim_shift}),
we employ phenomenological fits of the single-nucleon
inelastic structure functions. The latter 
are measured in DIS experiments and
a variety of parameterizations for $W_1$ and $W_2$ can be found in the
literature \cite{BodekRitchie,Bodek,Stein,NMC,Whitlow,MRST}, including
some variations arising from the different assumptions 
made for how to extract the neutron structure
functions from deuteron data. 
Unless stated otherwise, in the following we adopt 
the Bodek {\it et al.} fit
of~\cite{BodekRitchie,Bodek,Stein}, 
which describes both the deep inelastic and
resonance regions. 
For the QE contributions, we employ the form factor parameterization
of~\cite{Hohler}.
The sensitivity of the results to the 
different parameterization choices will be
discussed later.

Additionally, for the Fermi momentum and the energy
shift we will employ the values obtained in~\cite{Maieron:2001it}, namely
$k_F=220$ MeV/c, $\omega_{shift}=20$ MeV for carbon,
$k_F=236$ MeV/c, $\omega_{shift}=18$ MeV for aluminum,
$k_F=241$ MeV/c, $\omega_{shift}=23$ MeV for iron 
and $k_F=245$ MeV/c, $\omega_{shift}=25$ MeV for gold.
 
\subsection{Cross sections}
\label{subsec:cross}

In this section we present our results for the cross sections in the
RFG and ERFG models and compare them with the available experimental data,
\cite{Whitney:1974hr,Day:md,Arnold:1983mw,Arrington:1998ps}.

\begin{figure}[p]
\includegraphics{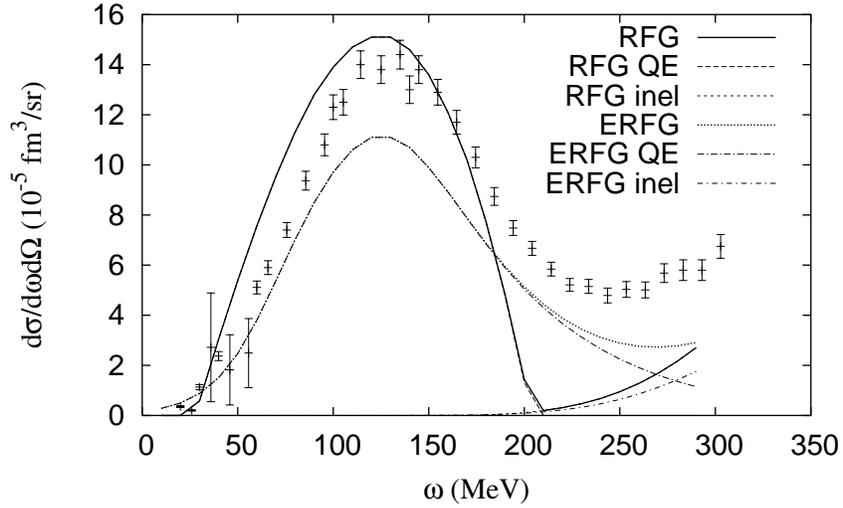}
\caption{Inclusive cross section for electron scattering
from carbon at $E_{inc}= 500$ MeV and $\theta_e=60^\circ$ versus
the energy transfer. The calculation includes an
energy shift $\omega_{shift}=20$ MeV and the separate
QE and inelastic contributions to the cross section are shown.
Data are from~\protect\cite{Whitney:1974hr}.}
\label{fig:sigHEPL_b}
\end{figure}

In Fig.~\ref{fig:sigHEPL_b} 
we show the inclusive cross section for a $^{12}$C
target at  $E_e=500$ MeV and $\theta_e=60^\circ$. 
We 
separate the QE from the inelastic contribution. 
We notice that the shifted RFG model (solid line) yields roughly the right 
position and height of the QE peak, but fails to reproduce 
the tails of the peak, 
giving in particular an unobserved dip at $\omega\simeq 200$ MeV. 
On the other hand the ERFG model (dotted line), 
while reproducing the data in
the tails better, significantly underestimates the cross section at the peak.
This is related to the fact that, as shown in Fig.~\ref{fig:fmodel},
the peak of the ERFG universal function $f_{ERFG}$ is lower than the
corresponding RFG value. 
Due to the larger extension of $f_{ERFG}$ over $\psi_X'$ the normalization
of the two functions is the same, namely
$\displaystyle{\int} f_{RFG} d\psi_X'=
\displaystyle{\int} f_{ERFG} d\psi_X'=1$. One might then naively
expect the integral in Eq.~(\ref{eq:rlt_funiv})
which yields the inelastic response functions to be the
same in the two models. However, a closer inspection shows 
that this is not the case
because the integration limits and/or the weighting provided by $U^{L,T}$
are such that the ERFG integral does not ``saturate" as does the RFG one.

\begin{figure}[p]
\includegraphics[width=0.8\textwidth]{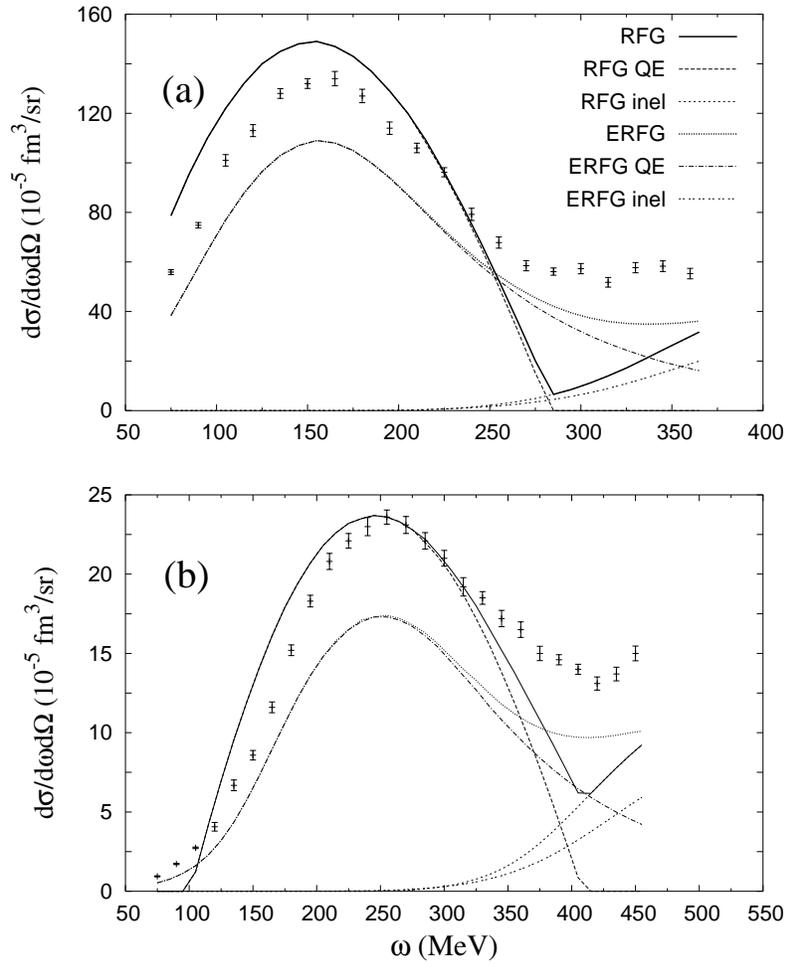}
\caption{As for Fig.~\ref{fig:sigHEPL_b}, but at
$E_{inc}= 2.020$ GeV and scattering angle $\theta_e=15^\circ$ (a) and
$\theta_e=20^\circ$ (b). 
Data are from \protect\cite{Day:md}.
}
\label{fig:CSLAC1}
\end{figure}

\begin{figure}[p]
\includegraphics[width=1.2\textwidth]{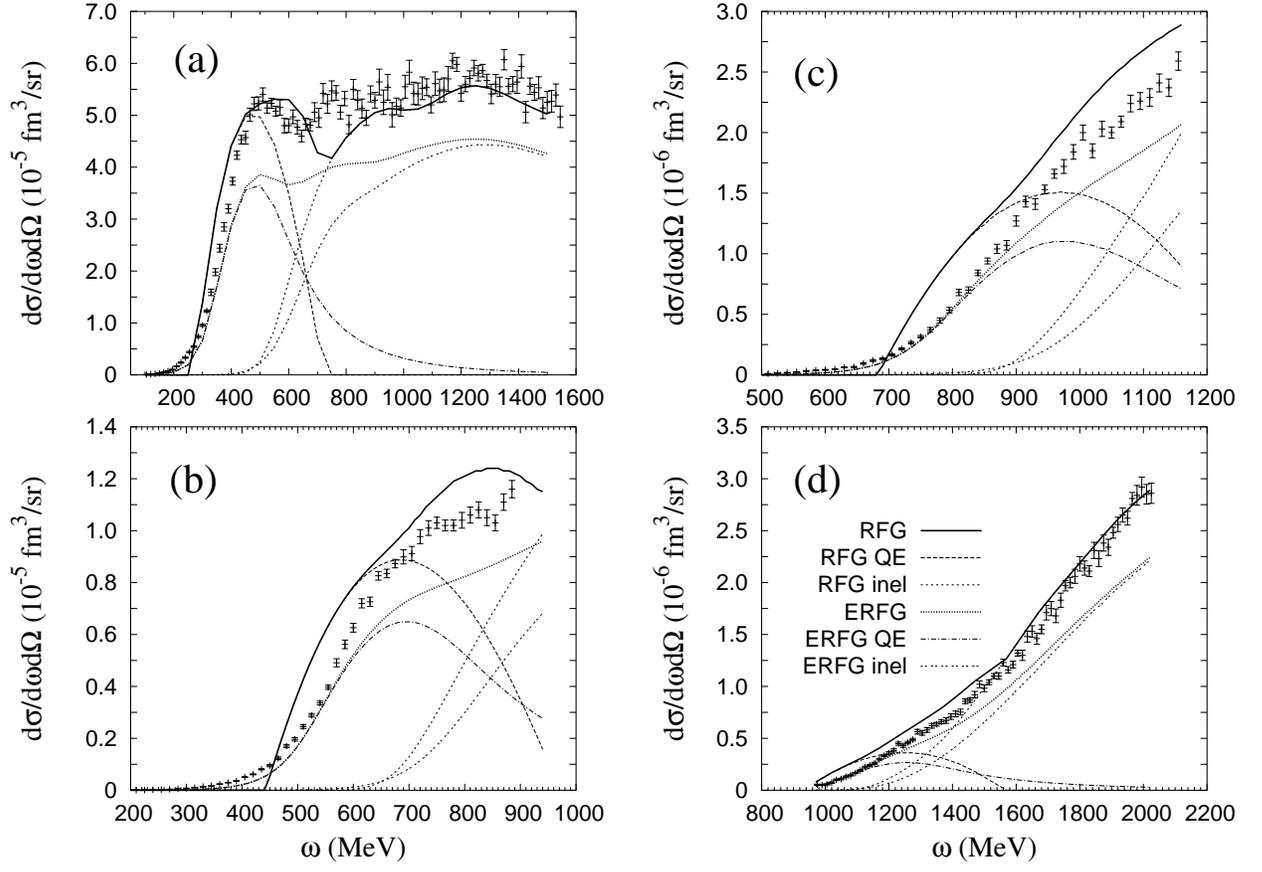}
\caption{As for Fig.~\ref{fig:sigHEPL_b}, but at $E_e=3.595$ GeV and 
scattering angle
$\theta_e=16^\circ$ (a), $20^\circ$ (b),
$25^\circ$ (c) and $30^\circ$ (d). 
Data are from \protect\cite{Day:md}.}
\label{fig:CSLAC2}
\end{figure}

\begin{figure}[p]
\includegraphics[width=1.2\textwidth]{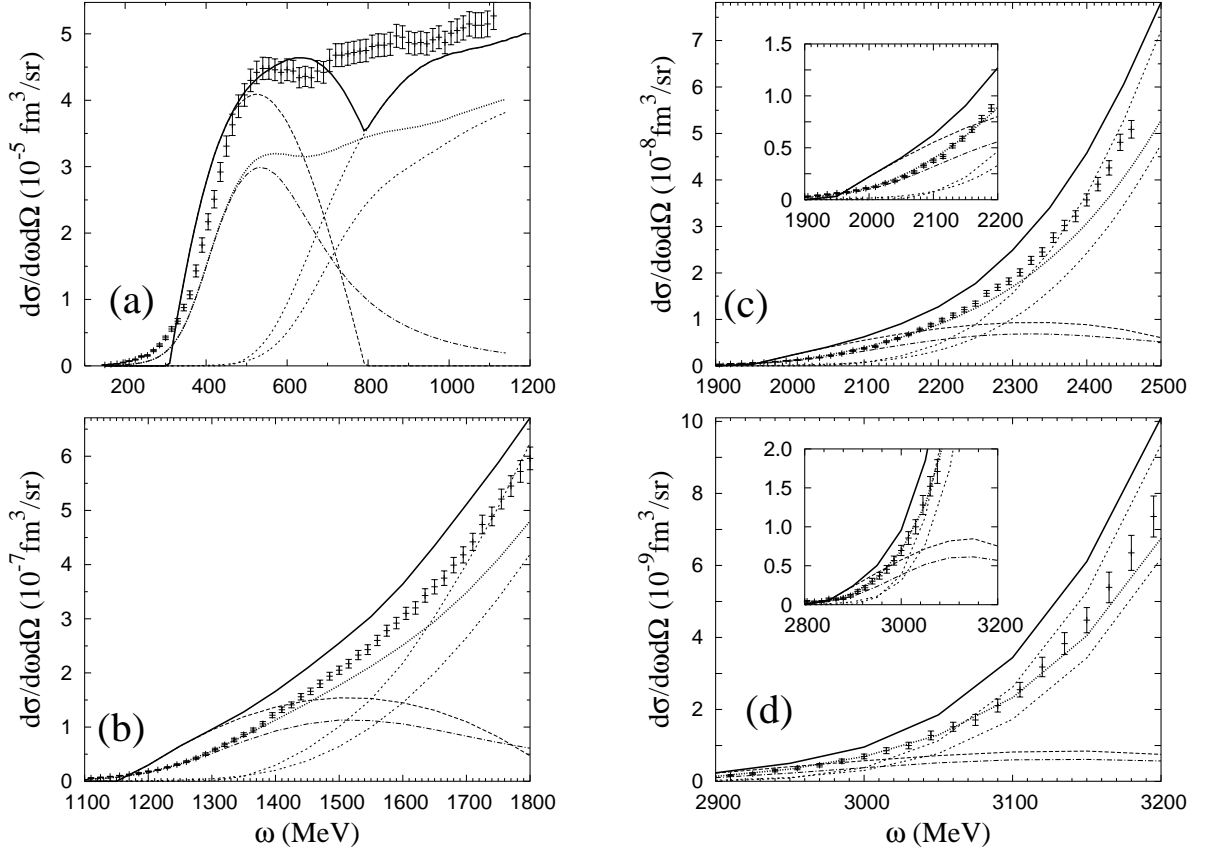}
\caption{As for Fig.~\ref{fig:sigHEPL_b}, but at
$E_{inc}= 4.045$ GeV and scattering angle 
$\theta_e=15^\circ$ (a), $30^\circ$ (b),
$45^\circ$ (c) and $74^\circ$ (d).
Data are from~\protect\cite{Arrington:1998ps}.}
\label{fig:CCEBAF}
\end{figure}

Figures~\ref{fig:CSLAC1}--\ref{fig:CCEBAF} 
correspond to different kinematical conditions, namely $E_e=2.020$ and $3.595$
GeV (SLAC) and $E_e = 4.045$ GeV (JLab) and various scattering angles.
Concerning Figs.~\ref{fig:CSLAC1}(a) and (b),
a similar trend persists, with the ERFG model 
significantly underestimating the data
in the region of the QE peak, whereas the RFG
is closer to the data (particularly for $\theta_e=20^\circ$), although it leaves
no room for other contributions to be added.
Note also that for this scattering angle the inelastic channel 
starts to be sizable.

Examining Figs.~\ref{fig:CSLAC2}  we remark that
the QE peak, which is more clearly separated from the inelastic region
in Fig.~\ref{fig:CSLAC2} (a), is again well reproduced in the 
low-$\omega$ tail by the ERFG, while its maximum agrees better with the RFG.
On the other hand the inelastic cross section is in all cases underestimated
by the ERFG, while the RFG alone would roughly 
account for what is observed.

Similar comments apply to Fig.~\ref{fig:CCEBAF} (a), 
corresponding to higher energy and low scattering angle.
For higher angles [Figs.~\ref{fig:CCEBAF} (b),(c),(d)] the data lie 
roughly in between the predictions of the ERFG (smaller) and RFG (larger)
models, the former again reproducing the low-$\omega$ behavior better.
As a general result we observe that as the scattering angle increases
the range of validity of the ERFG also increases.

\begin{figure}[p]
\includegraphics[width=0.9\textwidth]{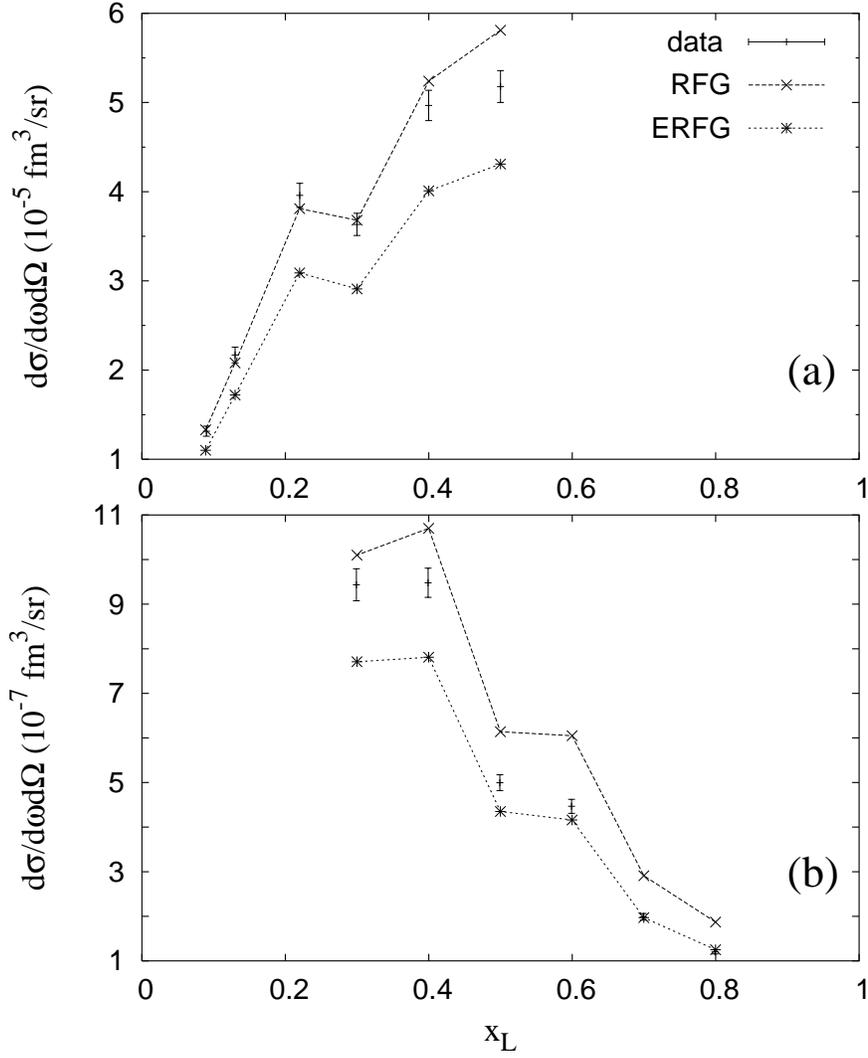}
\caption{Differential cross section $d\sigma/d\omega d\Omega_e$
for electron scattering on iron, shown as a function of $x_L$
at fixed $|Q^2|$.
Panel (a): $|Q^2| =2$ (GeV/c)$^2$, the experimental points, from right to left,
are taken at ($E_e$ (GeV),$\theta_e$ (degrees))= $(8,11.8),
(8,12.4),(8,13.6),(9.7,11.8),(12,11.8),(15,11.8) $.
Panel (b): $|Q^2| =10$ (GeV/c)$^2$, the experimental points, from right to left,
are taken at ($E_e$ (Gev),$\theta_e$ (degrees))= $(15,16.0),
(15,17.1),(17,15.0),(17,16.9),(21,14.1),(24,14.1) $.
Data are from \protect\cite{Arnold:1983mw}.
}
\label{fig:dis}
\end{figure}

\begin{figure}[p]
\includegraphics{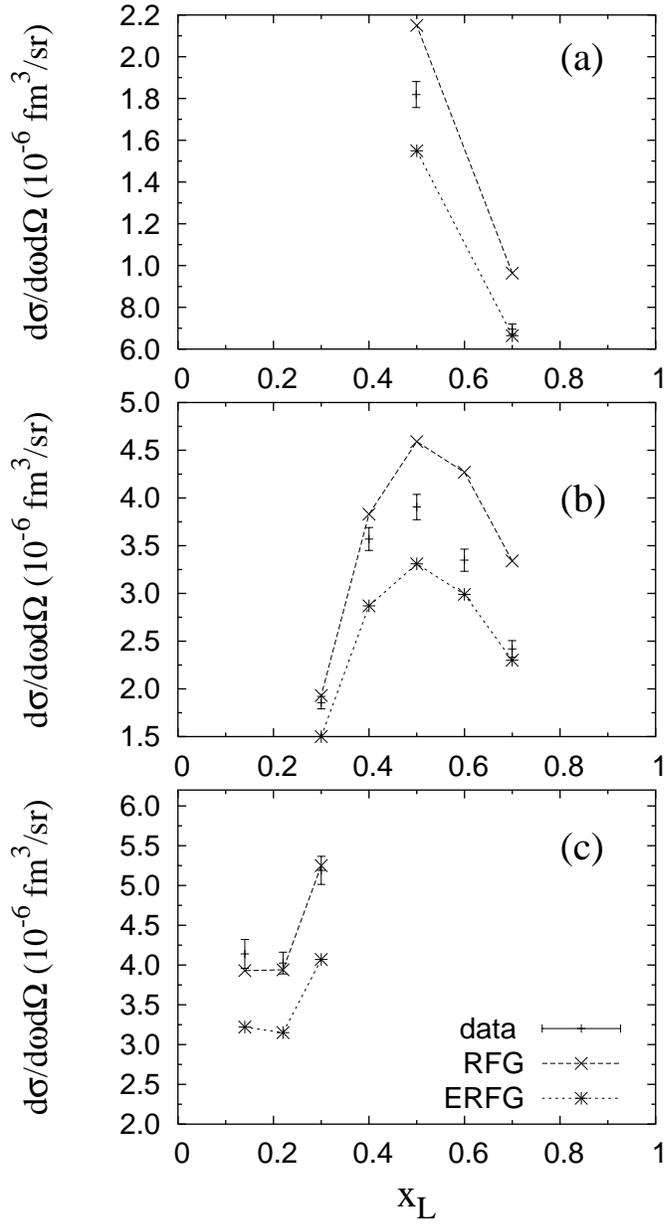}
\caption{As for Fig.~\ref{fig:dis}, but at $|Q^2|=5$ (GeV/c)$^2$.
The two data in the upper panel, from right to left,
correspond to 
$E_e=8$ 
GeV and 
$\theta_e=22^\circ$ 
and to $E_e=9.7$ GeV and $\theta_e=19.7^\circ$;
the data in the middle panel have fixed $E_e=12$ GeV and,
from right to left, $\theta_e= 12.8, 13.3, 14.2, 15.8,20.6^\circ$;
the data in the lower panel, from right to left, are taken at
($E_e= 15$ GeV, $\theta_e=13.2^\circ$), 
($E_e= 17$ GeV, $\theta_e=13.5^\circ$), 
($E_e= 24.5$ GeV, $\theta_e=11.1^\circ$).
Data are from \protect\cite{Arnold:1983mw}.
}
\label{fig:dis5}
\end{figure}

Finally, in Figs.~\ref{fig:dis} and \ref{fig:dis5}  we consider
slightly different kinematical conditions,
corresponding  to fixed values of $|Q^2|$ in the range 2-10 (GeV/c)$^2$,
and  various electron energies (in the range 8--25
GeV) and angles ($12^\circ$--$22^\circ$).
Theoretical results for $^{56}$Fe
are shown as functions of the 
``laboratory'' Bjorken variable $x_L$.  
The data corresponding to a fixed $Q^2$ are
taken at different values of $E_e$ and $\theta_e$.
For $|Q^2|=2$ and $10$ (GeV/c)$^2$ (Fig.~\ref{fig:dis})
the various data fit reasonably well on one plot, 
whereas for $|Q^2|=5$ (GeV/c)$^2$ (Fig.~\ref{fig:dis5}), for clarity
we have separated the data into three sets as indicated
in the figure caption.
We notice that at large $x_L$ ($\ge 0.6$) the data are closer to 
the ERFG predictions, at low $x_L$ (0.1-0.3) they are closer to the
RFG calculation and for $0.3\le x_L \le 0.6$ 
they lie in between the two models.
This general trend seems to be respected for all values of $Q^2$ (at least 
where data are available).
%
%

We have also analyzed the effect introduced by 
different electromagnetic form factor parameterizations 
(\cite{Galster,Iachello,Bosted})
and verified that it 
can produce a $\pm 3\%$ uncertainty at the QE peak, but does not
change the general agreement/disagreement of the models with the data.
Moreover, it should be remarked that, 
at the energies considered in this section,
the contribution from the resonance region 
to the inelastic part of the cross
section is quite important and thus a comparison with results 
obtained by using 
purely DIS parameterizations~\cite{NMC,MRST}
of the single-nucleon structure functions is not appropriate.
At the highest $|Q^2|$ values considered here [Figs.~\ref{fig:dis}(b),
\ref{fig:dis5}], the use of different parameterizations~\cite{NMC,MRST}
does not produce significant variations in the results.

An important comment, already anticipated in the Introduction, is 
in order. The RFG and ERFG models considered in this study include 
the 1p-1h one-body contributions both
for elastic scattering from a nucleon in the nucleus 
and for representations of
the single-nucleon inelastic spectrum, thereby incorporating 
effects from meson production,
excitation of baryon resonances (notably the $\Delta$) and, 
at high excitation energies,
DIS. However, this is not the entire story: in this
region and beyond effects arising from reaction mechanisms not 
included here, namely, those
coming from correlations and both 1p-1h and 2p-2h meson-exchange currents 
are also
important \cite{Barbaro:1995ez,Amaro:1998ta,Amaro:1999be,Amaro:2001xz,
Maieron:2001it,Amaro:2002mj,Amaro:2003yd,
DePace:2003xu}. 
In particular, in a recent study~\cite{DePace:2003xu}
effects from 2p-2h meson-exchange currents were explored for high-energy
conditions where relativistic modeling is important. 
The resulting cross sections are
significant in the region above the QE peak and therefore tend to bring the 
total (the present ERFG contributions plus these additional MEC contributions) 
into better agreement with the data. While this is encouraging, 
it is still not the full
story, since the 2p-2h MEC contributions have corresponding 
correlation contributions, as
required by gauge invariance (and as was studied in detail 
in our previous work on 1p-1h
MEC plus correlation effects \cite{Amaro:2003yd}). 
The 2p-2h correlations have not yet been
incorporated and thus detailed comparisons with data are somewhat premature.

In summary, the RFG model clearly overestimates the low-$\omega$ data, which 
are better reproduced by the ERFG model (dotted line), 
and the fact that the latter
yields a cross section that is below the data is encouraging, 
since this leaves room
for the above-mentioned effects to provide the balance. 

\subsection{Response and structure functions}
\label{subsec:resp}
In the RFG framework the only effect of the nuclear medium arises from
the Fermi motion of the nucleons inside the nucleus.  To quantify the
impact of the Fermi smearing on the observables we have compared the
inelastic RFG response functions with the corresponding ``unsmeared''
ones
\be R^{L,T}_{unsm.} \equiv Z R^{L,T}_{proton} + N
R^{L,T}_{neutron}\,,
\label{eq:rlt_unsm}
\ee
where $R^{L,T}_{proton\,(neutron)}$  
are the response functions for a free
proton (neutron) at rest in the laboratory frame
\footnote{The Bodek {\it et al.} fit we employ to describe the single-nucleon 
structure functions was obtained from data on cross sections
assuming a constant ratio $\sigma_L/\sigma_T$ = 0.18.  
When the fit is used to evaluate the separate
responses at relatively 
low $|Q^2|$, this may lead to some ``spurious'' effects,
such as the bump observed in the unsmeared $R_L$ at $\omega=0.86$ GeV, 
corresponding to the Delta resonance. This indicates that new and more 
precise fits of the nucleon structure functions in the resonance
region are needed.}.

A similar comparison has been done for the nuclear structure functions
$W_{1,2}^A$ and/or $F_{1,2}^A$ of Eqs.~(\ref{eq:W1A}--\ref{eq:F2A}).

In presenting the results we choose the following kinematical conditions: 
we select a relatively low (but typical)
four-momentum transfer $\tau=0.284$ (corresponding to $|Q^2|=1$ 
(GeV/c)$^2$), in order to illustrate the differences between
smeared and unsmeared quantities better.
The calculations are performed for the case of $^{56}$Fe with $k_F =
241$ MeV/c and they include
the energy shift discussed in Sec.~\ref{subsec:shift}, with $\omega_{shift}= 23$ 
MeV.

\begin{figure}[p]
\includegraphics[width=1.2\textwidth]{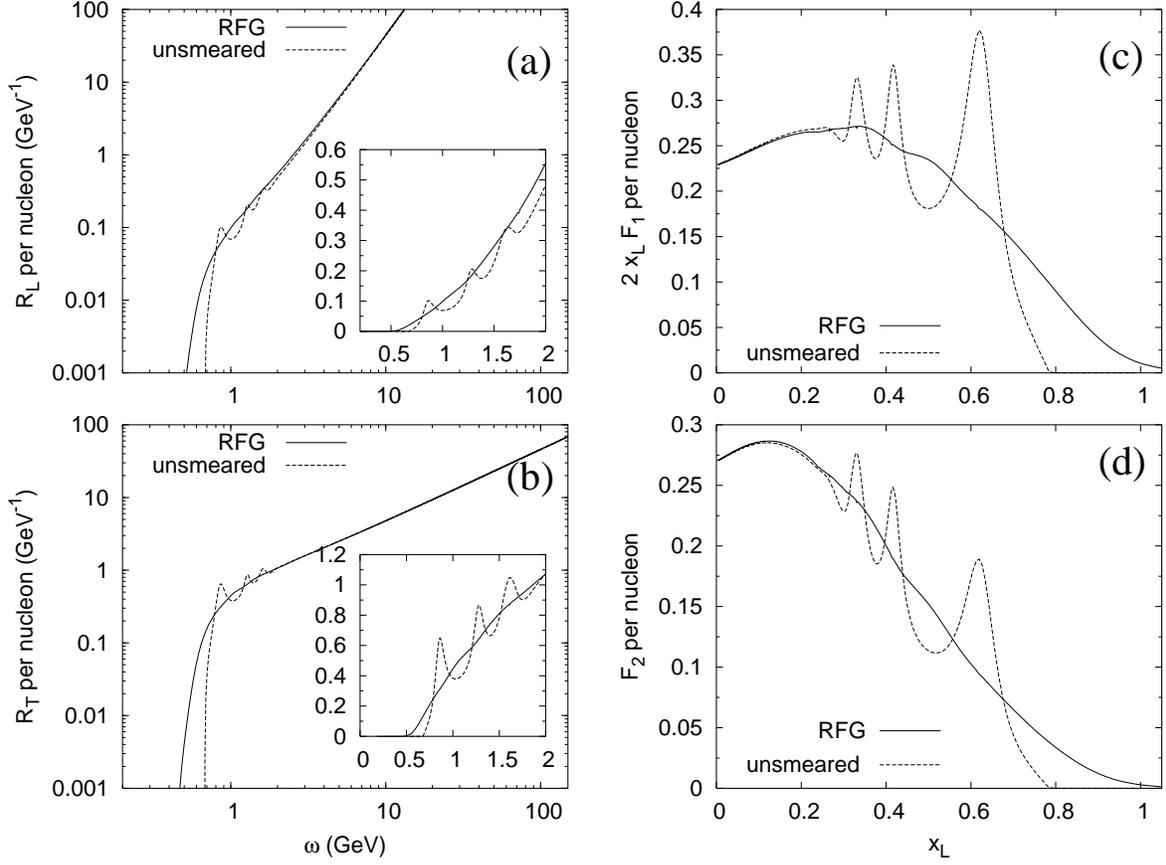}
\caption{
Panels (a) and (b):
inelastic and unsmeared response functions $R_{inel}^{L,T}$ per nucleon
as functions of the energy transfer $\omega$ for $^{56}$Fe,
calculated within the RFG, with $k_F=241$ MeV/c and 
$\omega_{shift}=23$ MeV. 
Panels (c) and (d):
inelastic and unsmeared structure functions 
$2 x_L F_1^A$ and $F_2^A$ per nucleon
as functions of $x_L = \tau/\lambda$.
}
\label{fig:rlt_smearing}
\end{figure}

In Fig.~\ref{fig:rlt_smearing}(a),(b) the inelastic response functions
$R_{inel}^{L,T}$ per nucleon are plotted as functions of the energy
transfer $\omega$,
while in Fig.~\ref{fig:rlt_smearing}(c),(d) the structure functions $2 x_L
F_1$ and $F_2$ are shown as functions of $x_L$.  
It is seen that in the resonance region è
[small panels in Fig.~\ref{fig:rlt_smearing}(a),(b)] the Fermi
smearing effects are rather large and completely smooth out the resonance 
structure of the single-nucleon responses, 
while in the DIS region (large $\omega$ or small
$x_L$) almost no effect of the nuclear medium is observed.

To illustrate the kind of effects introduced by the energy shift
accounted for following the procedure presented in
Appendix~\ref{app:shift}, we show in Fig.~\ref{fig:rlt_shift}(a) and (b)
the hadronic
responses, $R^{L,T}$, as functions of $\omega$.  The energy shift has
been taken to be $\omega_{shift}=23$ MeV. We present separately the
QE and inelastic channel contributions as well as the global
result. A similar analysis for the structure functions $2 x_L F_1^A$
and $F_2^A$, as given in Eqs.~(\ref{eq:F1A},\ref{eq:F2A}), is presented in
Fig.~\ref{fig:rlt_shift}(c) and (d). Note that the effects introduced by the
energy shift are observable in the QE peak and tend to 
disappear increasingly rapidly when moving to the inelastic region. In particular, it is
interesting to remark that the longitudinal response seems to
be more sensitive to inclusion of the energy shift. This is connected with the
large terms entering in ${\mathcal{D}}^\prime_L$ (see
Appendix~\ref{app:shift}). In this case the energy shift effects remain evident
even at very large $\omega$.

\subsection{Scaling functions}
\label{subsec:scaling}

In this section we investigate more closely the second-kind scaling 
behavior within the context of the inelastic RFG model. Since the
second-kind scaling analysis involves comparisons of different nuclear
species at the same kinematics and since a large ``reach'' in density
(or equivalently in Fermi momentum) is advantageous, we add the case of
gold to the discussions above.

In Fig.~\ref{fig:sigma} we plot the 
inclusive cross sections on gold for the
kinematical conditions $E_{inc}=3.6$ GeV, $\theta_e = 16^\circ$, and
compare them with available experimental data, taken at SLAC \cite{Day:md}.

Since it has been found to be desirable to have separate information
on longitudinal and transverse responses when discussing second-kind
scaling, we proceed as in past work \cite{Maieron:2001it} where
these data (for carbon, aluminum, iron and gold) were used 
to obtain ``L-subtracted'' transverse response functions and then transverse
superscaling functions. The L-subtraction was performed by assuming a
universal longitudinal superscaling function 
$f_L^{universal}(\psi^\prime)=f_{ERFG}(\psi^\prime)$ 
(see Sec.~\ref{subsec:ERFG})
and reconstructing from it the longitudinal cross section: \be
\Sigma_L = \frac{f_L^{universal}}{k_F} v_l {G}_L \sigma_M\,.
\label{eq:sigl}
\ee
$\Sigma_L$ was then subtracted from the total inclusive cross section
in order to obtain \be \Sigma_T = d\sigma -\Sigma_L \ee and then \ba
R^T &=& \frac{\Sigma_T}{\sigma_M v_T}
\label{eq:rt_lsub}
\\
f_T &=& \frac{k_F R^T}{{G}_T} \,.
\label{eq:ft_lsub}
\ea
In the above equations, according to~\cite{Maieron:2001it},
\ba
{G}_L &=& \frac{
\left(\kappa^2/\tau\right)
\left[\tilde{G}_E^2 + \tilde{W_2}{\Delta^\prime_T}\right]
}
{2 \kappa \left[1 + \xi_F\left(1 + \psi^{\prime^2}\right)/2\right]}
\label{eq:hatgl}
\\
{G}_T &=& \frac{ 2 \tau \tilde{G}_M^2 
+ \tilde{W_2}{\Delta^\prime_T}
}
{2 \kappa \left[1 + \xi_F\left(1 + \psi^{\prime^2}\right)/2\right]} 
\label{eq:hatgt}
\ea
with $\Delta^\prime_T$ defined in Eq.~(\ref{eq:delta_t}).

In Fig.~\ref{fig:rt} we show for the case of gold the ``L-subtracted'' (according to the
procedure described above) data for $R^T$ and compare them with the
theoretically calculated $R^T$, including both QE and inelastic
contributions.  Note that the transverse results
obtained via this subtraction procedure display a shortfall 
at high inelasticity of ``data'' versus inelastic RFG modeling, which
is not apparent in the total cross section shown in the previous figure.
This can be due to the fact that in subtracting the
longitudinal part, when elaborating the data, we may be using a 
longitudinal cross section that is too small, or, when assuming a
certain parameterization for the single-nucleon ratio
$W_2/W_1$ (related to $R$) to obtain 
the theoretical curves, we may be indirectly assuming a $\Sigma_L$ 
that is too large. Moreover, as discussed above, there is still a
2p-2h MEC plus correlation contribution to be taken into account
(note: the 2p-2h MEC contribution is predominantly transverse and so
this result is not unexpected).

Similar results are obtained for the superscaling functions
$f(\psi^\prime)$ and $f_T(\psi^\prime)$.
In Fig.~\ref{fig:effe} the total scaling function $f$ is shown as a
function of the QE variable $\psi^\prime$ for the four nuclear species
under discussion, within the RFG (left panel) and ERFG (right panel)
models, at the same kinematics of figs.~\ref{fig:sigma} and
\ref{fig:rt};
experimental data are obtained from the measured inclusive cross
sections divided by 
\be 
\sigma_M\left(v_L {G}_L + v_T {G}_T\right)
\label{eq:denom}
\ee 
and the curves are obtained by dividing the theoretical inclusive
cross section by the same quantity as in Eq.~(\ref{eq:denom}).

Finally, Fig.~\ref{fig:ft} (corresponding to Fig.~5 of~\cite{Maieron:2001it})
shows the transverse superscaling function $f_T(\psi^\prime)$, at the 
same kinematics, again
in the RFG and ERFG models.  The
``data'' are obtained from the experimental inclusive cross sections
according to Eqs.~(\ref{eq:sigl})--(\ref{eq:ft_lsub}), while the
curves are obtained by dividing the theoretical ${R^T}$ of Eqs.~(\ref{eq:totalresp}),
 (\ref{eq:rlt_qe}) and (\ref{eq:rlt_shift}) by
${G}_T$.  Again we observe that, 
the discrepancy between ``data'' and
``theory'' is larger for the transverse case than for the total
scaling functions at this scattering angle
($\theta=16^0$). This indicates that extra contributions should
be added to the nuclear model, going beyond the present one-body
description, and that these must act mainly in the transverse channel.
We have also checked that, in agreement with what previously observed,
when $\theta$ increases the difference
between total and transverse superscaling functions is more and more
reduced and thus
the disagreement between ``data'' and theory becomes the same
for $f$ and $f_T$.
 
When examining the last two figures we see that the basic trend in the
second-kind scaling behavior is present in the inelastic RFG modeling:
for fixed kinematics the heavier nuclei with the larger values of $k_F$
have the higher responses at high inelasticity, and by roughly the right
amount.

\section{Conclusions}
\label{sec:concl}
We have studied highly inelastic electron nucleus scattering, from the 
resonance to the DIS region, in a unified relativistic framework.
In particular we have calculated inclusive cross sections, response
functions and scaling functions in the Relativistic Fermi Gas and in a
phenomenological extension of it, named the Extended Relativistic Fermi Gas (ERFG), 
based on a fit of the scaling function in the quasielastic region.
We have explored all high quality experimental data available in the relevant
high-energy domain, 
involving energy transfers from zero up to
$\sim$ 3 GeV.

As discussed in detail in the results section the comparison between the data
and the theoretical models is strongly dependent upon the kinematics. However 
a few general features emerge from our analysis:
\begin{itemize}
\item
In the quasielastic regime the RFG model approximately accounts for the 
experimental strength of the peak, but fails to reproduce the 
low-$\omega$ tail of the cross sections and predicts a pronounced 
unobserved ``dip'' to the right of the QEP. On the contrary, the
ERFG model,
while correctly reproducing cross sections at low energy transfer,
always underestimates the data around the peak.
\item
In the highly inelastic part of the spectrum the RFG roughly yields the
experimental cross section for not too high energy transfer (corresponding
to smaller scattering angles) and overestimates the data when the inelasticity
becomes very high (large scattering angles). In parallel, the ERFG
underestimates the inelastic cross sections by $\sim$ 20\% at small $\theta$,
approaching  the data as $\theta$ ($\omega$) increases.  
\item
By analyzing the results in terms of the laboratory Bjorken scaling variable 
$x_L$ it is seen that the RFG works rather well at low $x_L$ (0.1-0.3),
whereas the ERFG is more appropriate to describe the high-$x_L$ ($\ge$ 0.6) 
data.
\item
A phenomenological energy shift is needed in both models to reproduce the QEP
position, but it is irrelevant in the highly inelastic region.
Moreover, concerning the separate responses,
the longitudinal one appears to be more sensitive to
the energy shift.
\item
The main impact of the nuclear medium on the responses and cross sections
consists in washing out the resonance structure present in the 
single-nucleon responses as a consequence of the 
Fermi motion of nucleons inside the
nucleus. In contrast, such an effect is negligible in the DIS regime.
\end{itemize}
The above findings point to the importance, 
in an intermediate region of energy
transfers, of ingredients 
which are not included in the present approach, such as meson-exchange 
currents and correlations, in both 1p-1h and 2p-2h sectors.
Preliminary results \cite{DePace:2003xu} seem to indicate that the
2p-2h MEC may play a crucial role in improving the agreement with the data,
although a complete and consistent calculation of  correlations 
and currents is still to be realized.
A separate analysis of the longitudinal and transverse response functions (or, 
equivalently, of the $F_1$ and $F_2$ structure functions) based on the scaling approach shows that these missing contributions should be mostly active in the
T channel, thus supporting the relevance of meson-exchange currents.

Finally, it is interesting to note that the disagreement between ERFG 
predictions and the experimental results is not peculiar to the specific
functional form of the phenomenological QE scaling function we have employed,
 but is 
essentially linked to its asymmetric shape. In fact, we checked that 
a simple ``toy model'' asymmetric 
scaling function (respecting of course the correct normalization) 
qualitatively yields similar results. We believe that the physical 
origin of this asymmetry is certainly worth further investigation.    

\acknowledgments
The authors wish to thank E. Barone for useful discussions on the DIS data,
M. Ripani for providing the references and programs
for the Bodek {\it et al.} 
fit of the inelastic structure functions and I. Sick for providing
data from experiments performed at SLAC and JLab.
This work was partially supported by DGI (Spain), 
under Contracts Nos BFM2002-03315, 
FPA2002-04181-C04-04, by the Junta de Andaluc\'{\i}a 
and by the INFN-CICYT exchange program.
The work of C.M. was supported in part by MEC (Spain) under
Contract SB2000-0427.
M.B.B. acknowledges financial support from MEC (Spain) for a
sabbatical stay at University of Sevilla (ref. SAB2001-0025), during which
part of this work was carried out. The work of T.W.D. was supported in part by funds 
provided by the U.S. Department of Energy under cooperative research agreement 
No. DE-FC02-94ER40818.

\appendix

\section{Inelastic tensor in RFG}
\label{app:RFG}

In this appendix we derive the general expression for 
$U^{\mu\nu}$ that enters in the hadronic inelastic
tensor in Eq.~(\ref{eq:Wadim4}).  By using Eqs.~(\ref{eq:Wadim1}) and
(\ref{eq:Wadim3}) we can write the hadronic tensor as follows:
\ba W^{\mu\nu}_{inel}(\kappa,\lambda) &=& \frac{3{\mathcal{N}}\tau}{2 \eta_F^3\kappa}
\int_{\rho_1(\kappa,\lambda)}^{\rho_2(\kappa,\lambda)}
d\rho \int_0^{2\pi}\frac{d\Phi}{2\pi}
\int_{\epsilon_0(\rho)}^{\epsilon_F} d{\overline\epsilon} \left[
  -w_1(\tau,\rho) \left( g^{\mu\nu}+\frac{\kappa^\mu\kappa^\nu}{\tau}
  \right) \right.
\nonumber \\
&+& \left.  w_2(\tau,\rho) \kappa^\mu\kappa^\nu\rho^2+ w_2(\tau,\rho)
  X^{\mu\nu} \right]\ ,
\label{eq:Ap1}
\ea having defined \ba X^{\mu\nu}&=& \eta^\mu\eta^\nu+\rho
\left( \eta^\mu\kappa^\nu+\eta^\nu\kappa^\mu \right) \ .
\label{eq:Xmunu}
\ea

To evaluate the above integral it is convenient to expand the
four-vector $\eta^\mu$ (which is normalized to $\eta_\mu\eta^\mu=1$)
in the basis $a^\mu=(\kappa,0,0,\lambda)$,
$\kappa^\mu=(\lambda,0,0,\kappa)$, $t_x^\mu=(0,1,0,0)$,
$t_y^\mu=(0,0,1,0)$, namely \be \eta^\mu = \eta_k\kappa^\mu+\eta_a
a^\mu+\eta_x t_x^\mu+\eta_y t_y^\mu
\label{eq:etamu}
\ee 
with \ba \eta_k &=& \eta\cos\theta_0 = -\rho
\label{eq:etak}
\\
\eta_a &=& \frac{1}{\kappa}\left(\epsilon+\lambda\rho\right)
\label{eq:etaa}
\\
\eta_x &=& \eta\sin\theta_0\cos\Phi
\label{eq:etax}
\\
\eta_y &=& \eta\sin\theta_0\sin\Phi \ .
\label{eq:etay}
\ea 
The integral of the tensor in Eq.~(\ref{eq:Xmunu}) then becomes 
\ba
&&\int_0^{2\pi}\frac{d\Phi}{2\pi} \int_{\epsilon_0(\rho)}^{\epsilon_F}
d{\overline\epsilon} X^{\mu\nu}= \int_0^{2\pi}\frac{d\Phi}{2\pi}
\int_{\epsilon_0(\rho)}^{\epsilon_F} d{\overline\epsilon} \left[
  \eta_k^2\kappa^\mu\kappa^\nu+\eta_a^2 a^\mu a^\nu +\eta_x^2 t_x^\mu
  t_x^\nu + \eta_y^2 t_y^\mu t_y^\nu \right.
\nonumber\\
&&+ \left.  \eta_k\eta_a\left(\kappa^\mu a^\nu+a^\mu\kappa^\nu\right)
  +\left(\eta_\kappa\kappa^\mu+\eta_a a^\mu\right)\kappa^\nu\rho
  +\kappa^\mu\rho \left(\eta_\kappa\kappa^\nu+\eta_a
    a^\nu\right) \right]
\nonumber\\
&&= \int_0^{2\pi}\frac{d\Phi}{2\pi} \int_{\epsilon_0(\rho)}^{\epsilon_F}
d{\overline\epsilon} \left[
  -\rho^2\kappa^\mu\kappa^\nu+\eta_a^2 a^\mu a^\nu +\eta_x^2
  t_x^\mu t_x^\nu + \eta_y^2 t_y^\mu t_y^\nu \right] \ ,
\ea 
since 
\be
\int_0^{2\pi} d\Phi \eta_x=\int_0^{2\pi} d\Phi\eta_y =\int_0^{2\pi}
d\Phi \eta_x\eta_y=0\ .  
\ee 
We now use the following integrals 
\ba
\int_0^{2\pi}\frac{d\Phi}{2\pi} \int_{\epsilon_0(\rho)}^{\epsilon_F}
d{\overline\epsilon} \eta_x^2 &=& \int_0^{2\pi}\frac{d\Phi}{2\pi}
\int_{\epsilon_0(\rho)}^{\epsilon_F} d{\overline\epsilon} \eta_y^2 =
\frac{1}{2} (\epsilon_F-\epsilon_0)
\theta\left(\epsilon_F-\epsilon_0\right) {\cal D}(\kappa,\tau,\rho)\ ,
\nonumber\\
\ea 
where ${\cal D}(\kappa,\tau,\rho)$ is given by Eq.~(\ref{eq:cald}), and
\ba \int_0^{2\pi}\frac{d\Phi}{2\pi} \int_{\epsilon_0(\rho)}^{\epsilon_F}
d{\overline\epsilon} \eta_a^2 &=& \int_0^{2\pi}\frac{d\Phi}{2\pi}
\int_{\epsilon_0(\rho)}^{\epsilon_F} d{\overline\epsilon} \frac{1}{\tau}
\left(1+\tau\eta_k^2+\eta_x^2+\eta_y^2\right)
\nonumber\\
&=& \left(\epsilon_F-\epsilon_0\right)
\theta\left(\epsilon_F-\epsilon_0\right) \left[1+\tau\rho^2+
  \frac{3}{2} {\cal D}(\kappa,\tau,\rho)\right] \ ,
\ea 
and observe that 
\be
t_x^\mu t_x^\nu +t_y^\mu t_y^\nu = - g^{\mu\nu} +\frac{a^\mu
  a^\nu}{\tau}-\frac{\kappa^\mu \kappa^\nu}{\tau} \ .  
\ee 
By inserting the above relations into Eq.~(\ref{eq:Ap1}) we get 
\ba
W^{\mu\nu}_{inel}(\kappa,\lambda) &=&\frac{3{\cal N}\tau}{2 \eta_F^3\kappa}
\int_{\rho_1(\kappa,\lambda)}^{\rho_2(\kappa,\lambda)}
d\rho
\left(\epsilon_F-\epsilon_0\right)\theta\left(\epsilon_F-\epsilon_0\right)
U^{\mu\nu}(\kappa,\tau,\rho)\ ,
\label{eq:Wmunufinal}
\ea 
with \ba U^{\mu\nu}(\kappa,\tau,\rho)&=&
-\left[w_1(\tau,\rho)+\frac{1}{2}w_2(\tau,\rho){\cal
    D}(\kappa,\tau,\rho)\right]
\left(g^{\mu\nu}+\frac{\kappa^\mu\kappa^\nu}{\tau}\right)
\nonumber\\
&+& w_2(\tau,\rho)\left[1+\tau\rho^2+\frac{3}{2}{\cal
    D}(\kappa,\tau,\rho)\right] \frac{a^\mu a^\nu}{\tau}\ .
\label{eq:Umunu}
\ea 
From the above expression the longitudinal and transverse
components in Eqs.~(\ref{eq:ul},\ref{eq:ut}) immediately follow.

The tensor in Eq.~(\ref{eq:Wmunufinal}) coincides with that in Eq.~(\ref{eq:Wadim4})
if the scaling variable $\psi_{_X}$ is introduced through the relation
\be \epsilon_F-\epsilon_0 = \xi_F\left(1-\psi_{_X}\right)^2\ .  \ee

\section{Inclusion of the energy shift}
\label{app:shift}

In this appendix we derive explicit expressions for the QE and inelastic 
hadronic
responses for the case in which a small energy shift is included in the 
analysis. As
mentioned in Sec.~4, to proceed one needs to assume a specific form for the
single-nucleon tensors and 
the variable dependence of the single-nucleon structure
functions. This choice is not unique and hence, some ambiguities enter in the
analysis of the results. 
Here we adopted a specific strategy (see below); however, the cautionary statement should be made that other choices 
are possible and that these can lead to different results for the observables.

The specific form of the single-nucleon tensor we have selected is
\be
w_{inel}^{\mu\nu} = -w_1(\tau,\tilde{\rho})
\left(
g^{\mu\nu} + \frac{\kappa^\mu \kappa^\nu}{\tau}
\right) 
+ w_2(\tau,\tilde{\rho})
\left(
\eta^\mu + \frac{\eta\cdot \kappa}{\tau}\kappa^\mu
\right)
\left(
\eta^\nu + \frac{\eta\cdot \kappa}{\tau}\kappa^\nu
\right) \, .
\ee
The longitudinal and transverse hadronic functions $U^{L,T}$ that result are
\ba
U^L&=&\frac{\kappa^2}{\tau}
\left[\left(1+\tau\rho^{\prime^2}\right)
w_2(\tau,\tilde{\rho})-
w_1(\tau,\tilde{\rho})
+w_2(\tau,\tilde{\rho}){\mathcal D}^\prime_L(\kappa,\tau,\lambda_{shift},
\rho^\prime)\right]
\label{eq:ulp}
\\
U^T&=&
2 w_1(\tau,\tilde{\rho})
+w_2(\tau,\tilde{\rho}){\mathcal D}^\prime_T(\kappa,\tau,\lambda_{shift},\rho^\prime)\ ,
\label{eq:utp}
\ea
where
\ba
{\mathcal D}^\prime_T(\kappa,\tau,\lambda_{shift},
\rho^\prime) &=& \xi_F(1-\psi_{_X}^{\prime^2})
\left[\frac{1}
{\kappa}\sqrt{\tau^\prime\left(1+\tau^\prime\rho^{\prime 2}\right)}
+\frac{1}{3}\xi_F\frac{\tau^\prime}{\kappa^2}(1+\psi_{_X}^{\prime^2})\right]
\label{eq:cald_t}\\
{\mathcal D}^\prime_L(\kappa,\tau,\lambda_{shift},
\rho^\prime) &=& \frac{\tau}{\kappa^2}
\left[
\left(
\frac{\lambda\tau^\prime\rho^\prime}{\tau}
+1 + \frac{\lambda}{\tau}\lambda_{shift}
\right)^2 
\right.
\nonumber \\ 
&+&
\left(
\frac{\lambda\tau^\prime\rho^\prime}{\tau}
+1 + \frac{\lambda}{\tau}\lambda_{shift}
\right)
\left(
1 + \frac{\lambda}{\tau}\lambda_{shift}
\right)
\xi_F(1+\psi_{_X}^{\prime^2}) 
\nonumber \\ 
&+&
\left.
\left(
1 + \frac{\lambda}{\tau}\lambda_{shift}
\right)^2
\frac{1}{3}\xi_F^2
(1+\psi_{_X}^{\prime^2}+\psi_{_X}^{\prime^4})
\right]
-\left(
1+\tau\rho^{\prime^2}
\right) \ .
\label{eq:cald_l}
\ea
Notice that terms like 
$\displaystyle{\frac{\lambda}{\tau}\lambda_{shift}}$ and
$\displaystyle{\frac{\tau^\prime}{\tau} =
1- \displaystyle{\frac{\lambda_{shift}^2}{\tau}}
+ 2\displaystyle{\frac{\lambda}{\tau}\lambda_{shift}}}$,
appearing in ${\mathcal D}^\prime_L$, can become large
when $x_L \equiv\displaystyle{\frac{\lambda}{\tau}}$ is small,
even if $\lambda_{shift}$ is small. We repeat that the above choice for the
single-nucleon tensor is not unique, and that other choices involving the four-momentum
$Q^{\prime\mu}$ instead of 
$Q^\mu$ are possible. Moreover, the arguments of the 
single-nucleon inelastic structure functions 
$w_{1,2}$, should be also considered
carefully. 
As discussed in Sec.~4, here the 
inelasticity parameter,
denoted as
$\tilde{\rho}$, is given by
\be
\tilde{\rho} \equiv \frac {2 H\cdot Q^\prime}{|Q^2|} =
\frac{W_{_X}^{\prime^2}-m_N^2 + |Q^{\prime^2}|}{|Q^2|}=
 \rho^\prime \frac{\tau^\prime}{\tau} \, 
\label{eq:xprime}
\ee
which coincides with the expression given as
$(2m_N\tilde{\omega})/|Q^2|$, with 
$\tilde{\omega}$ being the energy transferred to the
nucleon in the system in which the nucleon is at rest. 
One should be aware that other alternatives exist,
in particular, one can consider the inelasticity $\rho$
corresponding to a free nucleon at rest with final-state 
invariant mass equal to $W_{_X}'$. 

The ambiguity introduced in the inelastic responses due to the energy shift 
is also present
at the level of the hadronic QE response functions. Again, the problem is 
directly connected
with the form assumed for the single-nucleon tensor. For consistency with 
the formalism used
in the inelastic channel, in the QE process 
the single-nucleon tensor $w^{\mu\nu}_{QE}$ is taken to be
\be
w_{QE}^{\mu\nu} = -w_{1,QE}(\tau)
\left(
g^{\mu\nu} + \frac{\kappa^\mu \kappa^\nu}{\tau}
\right) 
+ w_{2,QE}(\tau)
\left(
\eta^\mu + \frac{\eta\cdot \kappa}{\tau}\kappa^\mu
\right)
\left(
\eta^\nu + \frac{\eta\cdot \kappa}{\tau}\kappa^\nu
\right) \, .
\ee
The hadronic QE response functions within the RFG model are given by
\be
R_{QE}^{L,T} =
\frac{3{\mathcal{N}}}
{4 \eta_F^3\kappa m_N} 
\xi_F \left(1-\psi^{\prime^2}\right) 
\theta(1-\psi^{\prime^2})U_{QE}^{L,T} 
\label{eq:rlt_qe} 
\ee
with the structure functions 
\ba
U_{QE}^L &=& \frac{\kappa^2}{\tau}
\left[ (1+ \tau)w_{2,el}(\tau) - w_{1,el}(\tau)
+ w_{2,el}(\tau) \Delta^\prime_L(\tau,\kappa,\lambda_{shift})
\right]
\label{eq:ul_qe}
\\
U_{QE}^T &=& 2 w_{1,el}(\tau) + w_{2,el}(\tau) 
\Delta^\prime_T(\tau,\kappa,\lambda_{shift})
\label{eq:ut_qe} \, .
\ea
The nuclear structure dependence is contained in 
the terms $\Delta^\prime_{L,T}$ in the form
\ba
\Delta^\prime_T(\kappa,\tau,\lambda_{shift})
&=& \xi_F(1-\psi^{\prime^2})
\left[\frac{1}{\kappa}\sqrt{\tau^\prime\left(1+\tau^\prime\right)}
+\frac{1}{3}\xi_F\frac{\tau^\prime}{\kappa^2}(1+\psi^{\prime^2})\right]
\label{eq:delta_t}
\\
\Delta^\prime_L(\kappa,\tau,\lambda_{shift})
&=& \frac{\tau}{\kappa^2}
\left[
\left(
\lambda\frac{\tau^\prime}{\tau}
+1 + \frac{\lambda}{\tau}\lambda_{shift}
\right)^2 
\right.
\nonumber \\ 
&+&
\left(
\lambda\frac{\tau^\prime}{\tau}
+1 + \frac{\lambda}{\tau}\lambda_{shift}
\right)
\left(
1 + \frac{\lambda}{\tau}\lambda_{shift}
\right)
\xi_F(1+\psi^{\prime^2}) 
\nonumber \\ 
&+&
\left.
\left(
1 + \frac{\lambda}{\tau}\lambda_{shift}
\right)^2
\frac{1}{3}\xi_F^2
(1+\psi^{\prime^2}+\psi^{\prime^4})
\right]
-\left(
1+\tau
\right) \, .
\label{eq:delta_l}
\ea
The scaling variable $\psi^{\prime}$ is given by
\be
\psi^{\prime^2} = \frac{1}{\xi_F}
\left( \kappa \sqrt{\frac{1}{\tau^\prime} +1}
- \lambda^\prime -1
\right) \,
\label{eq:psiqe}
\ee
and the electromagnetic structure functions by
\ba
w_{1,QE}(\tau) &=& \tau G_M^2(\tau) 
\nonumber \\
w_{2,QE}(\tau) &=& 
\frac{G_E^2(\tau) + \tau G_M^2(\tau)}{1 + \tau} 
\label{eq:w12_el}
\ea
with $G_{E,M}$ the proton or neutron Sachs electromagnetic form factors.

Finally, the ``total'' response functions are evaluated by adding the above QE responses to
the inelastic ones, i.e., $R^{L,T}_{tot}=R^{L,T}_{QE}+R^{L,T}_{inel}$.


\newpage
\newpage
\newpage
\newpage

%
\newpage

\newpage
\begin{figure}[ht]
\includegraphics[width=1.2\textwidth]{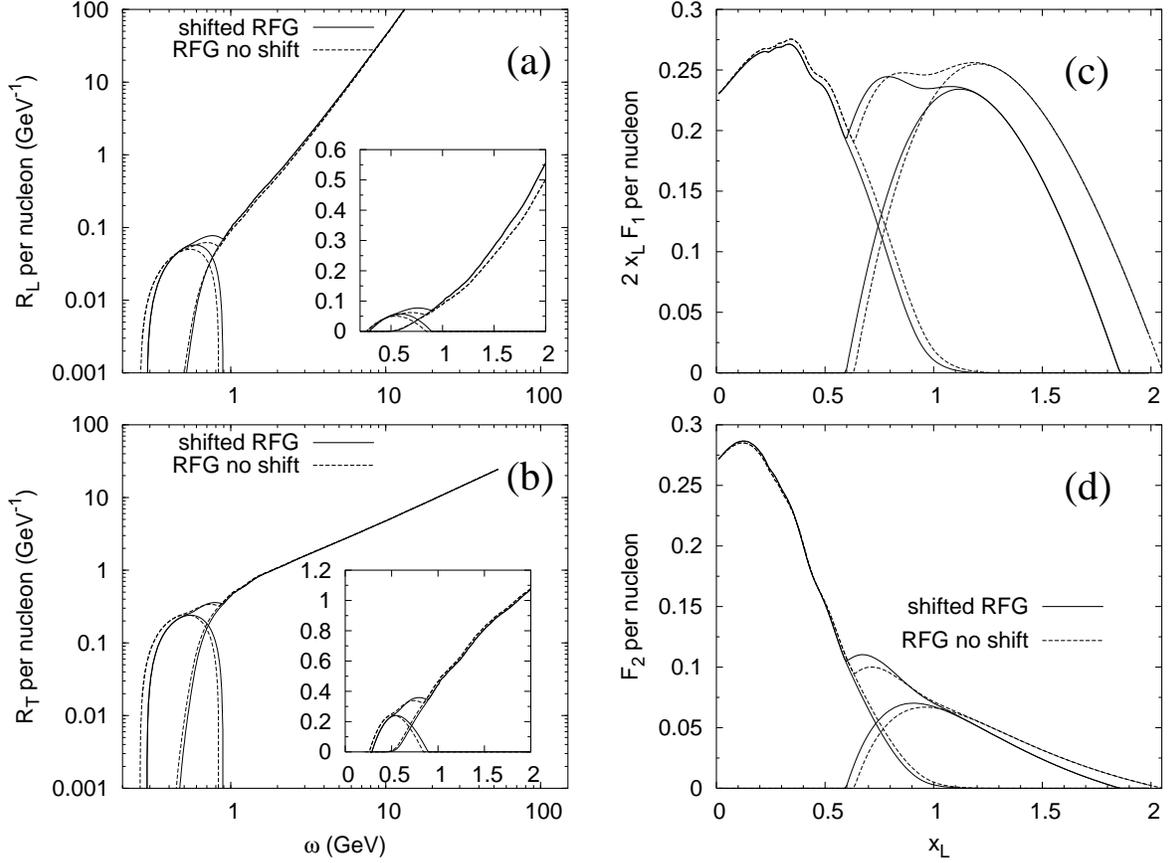}
\caption{
Panels (a) and (b):
response functions $R_L$ and $R_T$
per nucleon, respectively, as functions of $\omega$ for $^{56}$Fe.
The separate contributions of the QE peak
(visible at low $\omega$) and of the inelastic response functions 
are shown, together with the sum of the two contributions.
Panels (c) and (d): structure functions $2 x_L F_1$ and, respectively
$F_2$ per nucleon versus $x_L$. 
In this case the contribution of the QE peak is the
one on the right (large $x_L$) and, correspondingly, the inelastic
contribution is the structure appearing at small $x_L$.
}
\label{fig:rlt_shift}
\end{figure}

\newpage
\begin{figure}[p]
\includegraphics{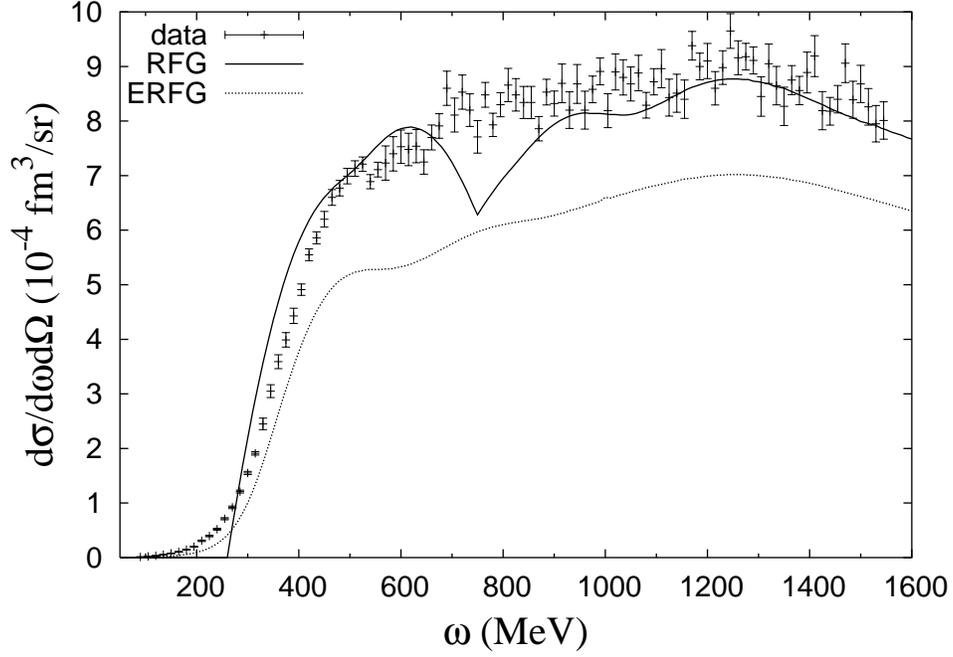}
\caption{Inclusive cross section for electron scattering
from gold, at $E_{inc}= 3.6$ GeV and $\theta_e=16^\circ$ versus
the energy transfer. Data are from  \protect\cite{Day:md}.
}
\label{fig:sigma}
\end{figure}
\newpage
\begin{figure}[p]
\includegraphics{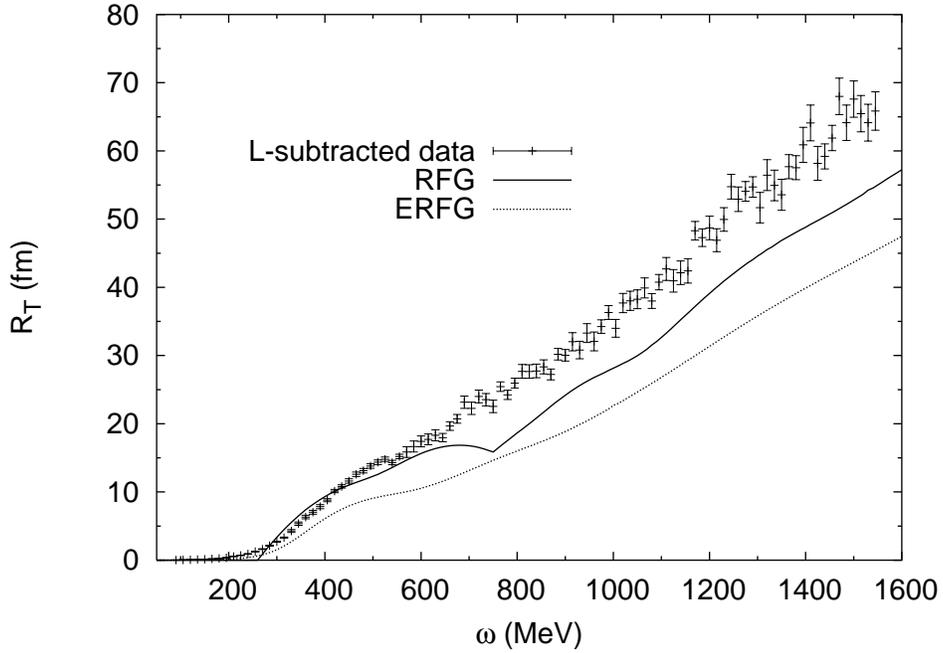}
\caption{Transverse response function for electron scattering
from gold, at $E_{inc}= 3.6$ GeV and $\theta_e=16^\circ$ versus
the energy transfer. The ``data'' are obtained from measured inclusive
cross sections by means of the L-subtraction 
procedure described in the text.
}
\label{fig:rt}
\end{figure}
\newpage

\begin{figure}[p]
\includegraphics[width=1.1\textwidth]{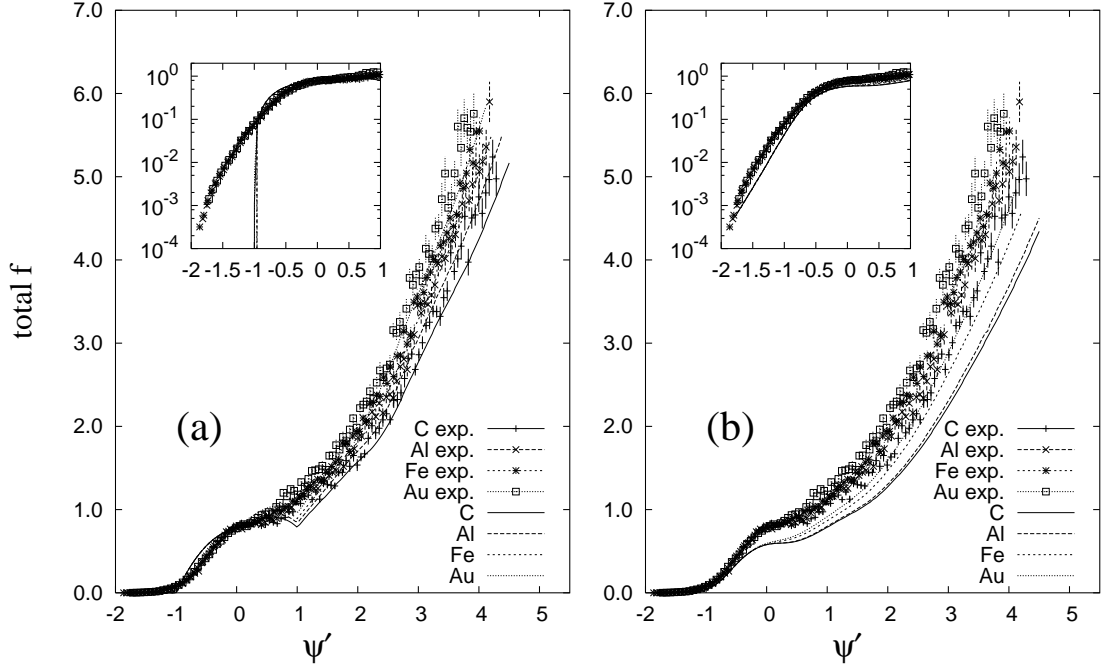}
\caption{Total superscaling functions $f(\psi^\prime)$, as described
in Sec.~\ref{subsec:scaling}, for the kinematical conditions
$E_e=3.595$ GeV and $\theta_e=16^\circ$. Theoretical results obtained
within the RFG are shown in panel (a), while the ERFG case is presented
in panel (b).
}
\label{fig:effe}
\end{figure}
\newpage
\begin{figure}[p]
\includegraphics[width=1.1\textwidth]{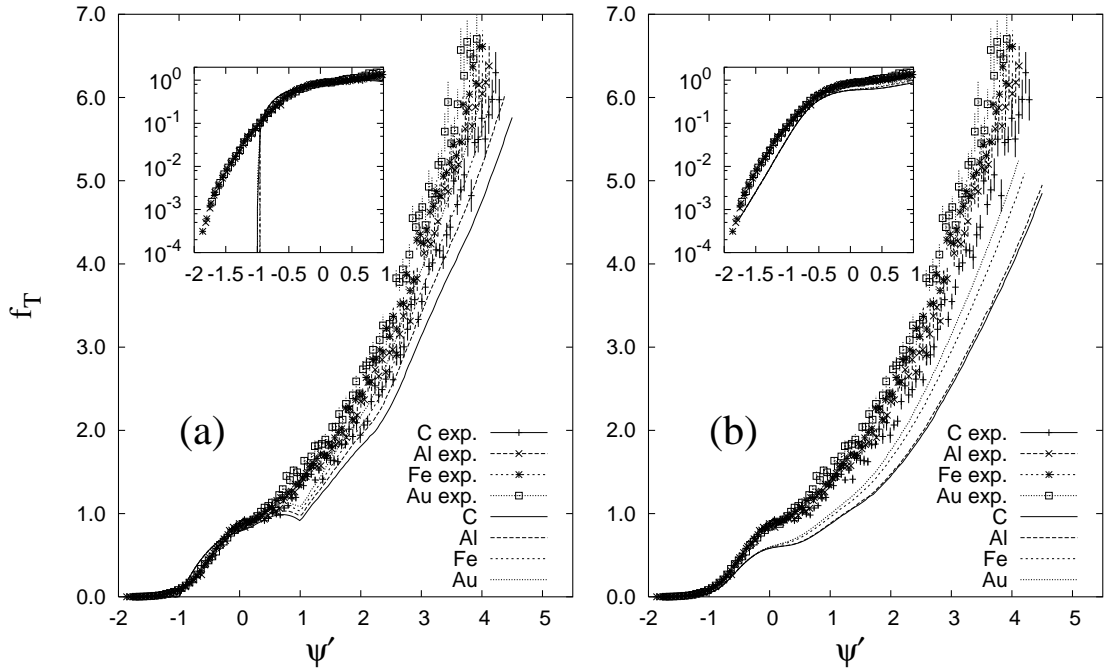}
\caption{As for Fig.\ref{fig:effe}, but showing the transverse superscaling
functions $f_T(\psi^\prime)$ (see discussion in
Sec.~\ref{subsec:scaling}).
}
\label{fig:ft}
\end{figure}

\end{document}